\documentclass[%
 reprint,
 superscriptaddress,
 nofootinbib, 
 nobibnotes,
 amsmath,amssymb,
 aps,
pdftex,
]{revtex4-1}

\usepackage{graphicx}
\usepackage{dcolumn}
\usepackage[hypertexnames=false]{hyperref}
\usepackage{enumerate}
\usepackage{bm}
\usepackage{mathrsfs}
\usepackage{comment}
\usepackage{url}
\usepackage{siunitx}
\usepackage[nolist]{acronym}
\usepackage{color}

\usepackage[utf8x]{inputenc}
\usepackage[T1]{fontenc}

\newcommand{\unit}[1]{\,\rm{#1}}
\newcommand{\eq}[2][]{
	\begin{align#1}
		#2
	\end{align#1}
}

\newcommand{\myfig}[4][width=\linewidth]{
    \begin{figure}[tbp]
        \centering
            \includegraphics[#1]{#2}    
        \caption{#3}                    
        \label{#4} 
    \end{figure}
}

\newcommand{\diff}{\mathrm{d}}
\newcommand{\iu}{\mathrm{i}}
\newcommand{\e}{\mathrm{e}}

\newcommand{\cross}{\times}
\newcommand{\figref}[1]{Fig.~\ref{#1}}
\newcommand{\tabref}[1]{Table~\ref{#1}}
\newcommand{\secref}[1]{Sec.~\ref{#1}}
\newcommand{\appref}[1]{Appendix~\ref{#1}}

\allowdisplaybreaks[4] 

\definecolor{mygreen}{RGB}{0,130,0} 

\begin{document}


\title{Early warning of precessing compact binary merger with third-generation gravitational-wave detectors}

\author{Takuya Tsutsui}
\affiliation{%
Research Center for the Early Universe (RESCEU), Graduate School of Science, The University of Tokyo, Tokyo 113-0033, Japan
}
\affiliation{%
Department of Physics, Graduate School of Science, The University of Tokyo, Tokyo 113-0033, Japan
}
\author{Atsushi Nishizawa}%
\affiliation{%
Research Center for the Early Universe (RESCEU), Graduate School of Science, The University of Tokyo, Tokyo 113-0033, Japan
}
\author{Soichiro Morisaki}%
\affiliation{%
Department of Physics, University of Wisconsin-Milwaukee, Milwaukee, WI 53201, USA
}

\date{\today}

\begin{abstract}
Rapid localization of gravitational-wave events is important for the success of the multi-messenger observations.
The forthcoming improvements and constructions of gravitational-wave detectors will enable detecting and localizing compact-binary coalescence events even before mergers, which is called early warning.
The performance of early warning can be improved by considering modulation of gravitational wave signal amplitude due to the Earth rotation and the precession of a binary orbital plane caused by the misaligned spins of compact objects.
In this paper, for the first time we estimate localization precision in the early warning quantitatively, taking into account an orbital precession.
We find that a neutron star-black hole binary at $z=0.1$ can typically be localized to $100\unit{deg}^2$ and $\SI{10}{deg^2}$ at the time of $12$ -- $15 \unit{minutes}$ and $50$ -- $\SI{300}{seconds}$ before merger, respectively, which cannot be achieved without the precession effect.

\end{abstract}

\maketitle


\section{Introduction} \label{SEC:introduction}
The first joint detection of a gravitational wave (GW) and electromagnetic (EM) radiation in broad bands from binary neutron-stars (BNS) coalescence has opened a new era of multi-messenger astronomy~\cite{GW170817_observation, GW170817_multimessenger}, which is collaborations with EM and neutrino observations.
Currently the information about event triggers are shared as open public alerts, which had been used for EM follow-ups in the third observing run (O3).
So far, GWs from binaries have been detected after the mergers.
However, if the signals are sufficiently large, those can be detected before the mergers by accumulating enough signal-to-noise ratio (SNR), which is called early warning~\cite{early_warning, early_warning_simulation}.
Early warning should be important for a success of the follow-up observations and obtaining information on precursor events, prompt emission from BNS merger~\cite{prompt_flash}, characteristic EM emission from tidal disruption of neutron star-black hole (NSBH)~\cite{NSBH_EMemission}, resonant shattering~\cite{resonant_shattering}, fast radio burst driven by black-hole (BH) battery~\cite{FRB_BH_battery}, and so on.

Until the former half of the O3, many GW events have been detected~\cite{GWTC-1, GWTC-2} by advanced LIGO~\cite{LIGO1, LIGO2} and advanced Virgo~\cite{Virgo}.
In 2030s, constructions of third-generation (3G) GW detectors, Einstein Telescope (ET)~\cite{ET_paper, ET_science} and Cosmic Explorer (CE)~\cite{CE1, CE2} are planned.
The 3G GW detectors are much more sensitive than advanced LIGO (see \figref{FIG:PSD}), and then GWs can be observed from the frequency as low as $\sim \SI{1}{Hz}$.
\myfig[width=0.9\linewidth]{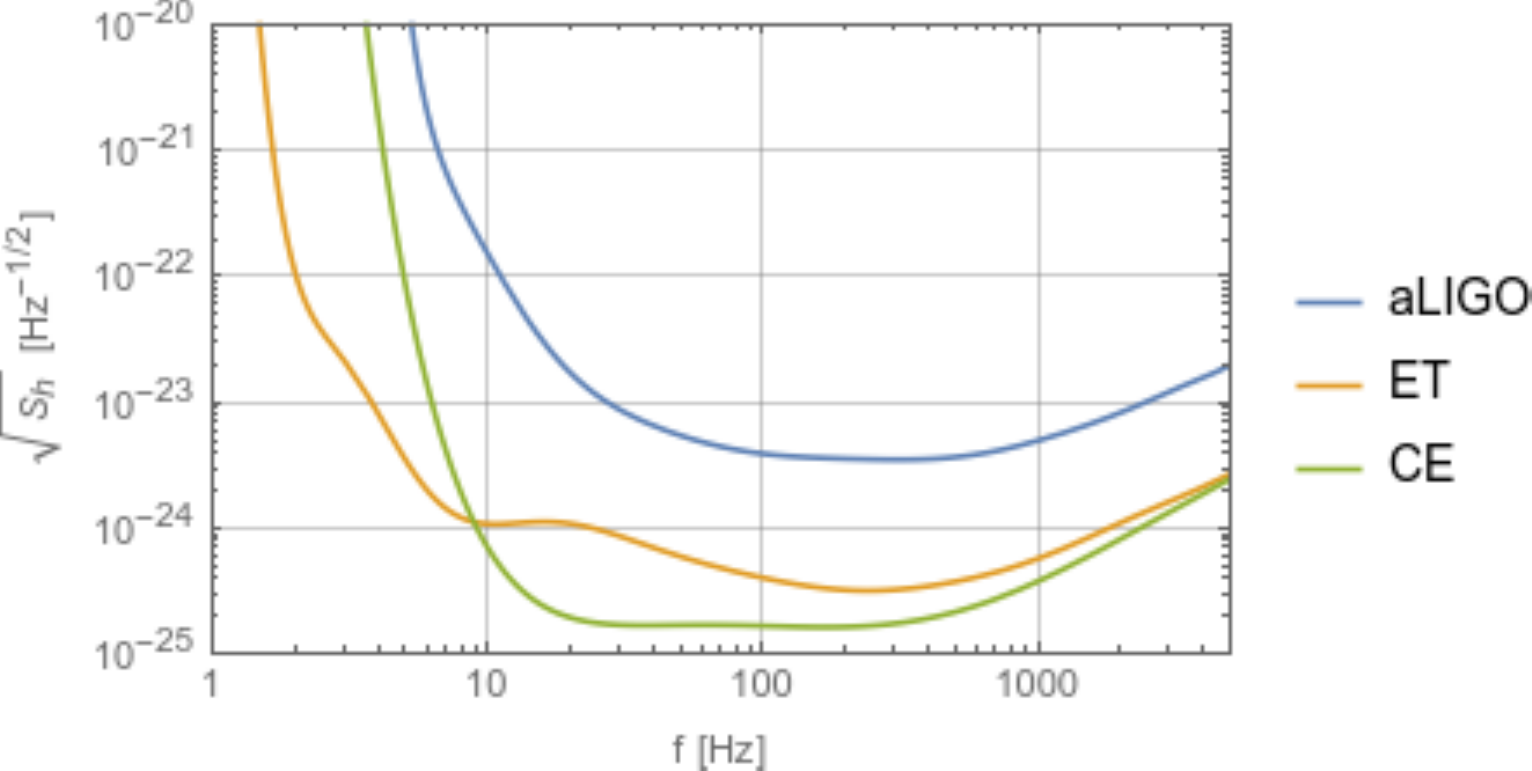}{Power spectral densities for advanced LIGO (blue), ET-D (orange) and CE (green).}{FIG:PSD}
Accordingly, the time to merger of binaries of, for example, neutron stars (NSs) with $\SI{1.4}{M_\odot}$ and {BH} with $\SI{14}{M_\odot}$ is $\sim \SI{1}{day}$.
Since the Earth rotates at the period of one day, GWs from the binaries are modulated through the antenna response functions.
In addition, the Doppler phases vary in time. These effects improve localization of the binaries at lower frequencies~\cite{earlywarning_3G, ET_CE_BNS_doppler, earlywarning_BNS} and help early warning.
For further improvement of the early warning, the inclusion of higher order modes in a signal analysis has been suggested~\cite{earlywarning_highermode}. 

In this paper, we investigate the performance of the early warning by {3G} {GW} detectors when precession effects of compact binaries are considered. If either or both of compact objects have non-zero misaligned spins, the total angular momentum of a binary is not parallel with the orbital angular momentum (see \figref{FIG:concept_precession}).
\myfig[width=0.6\linewidth]{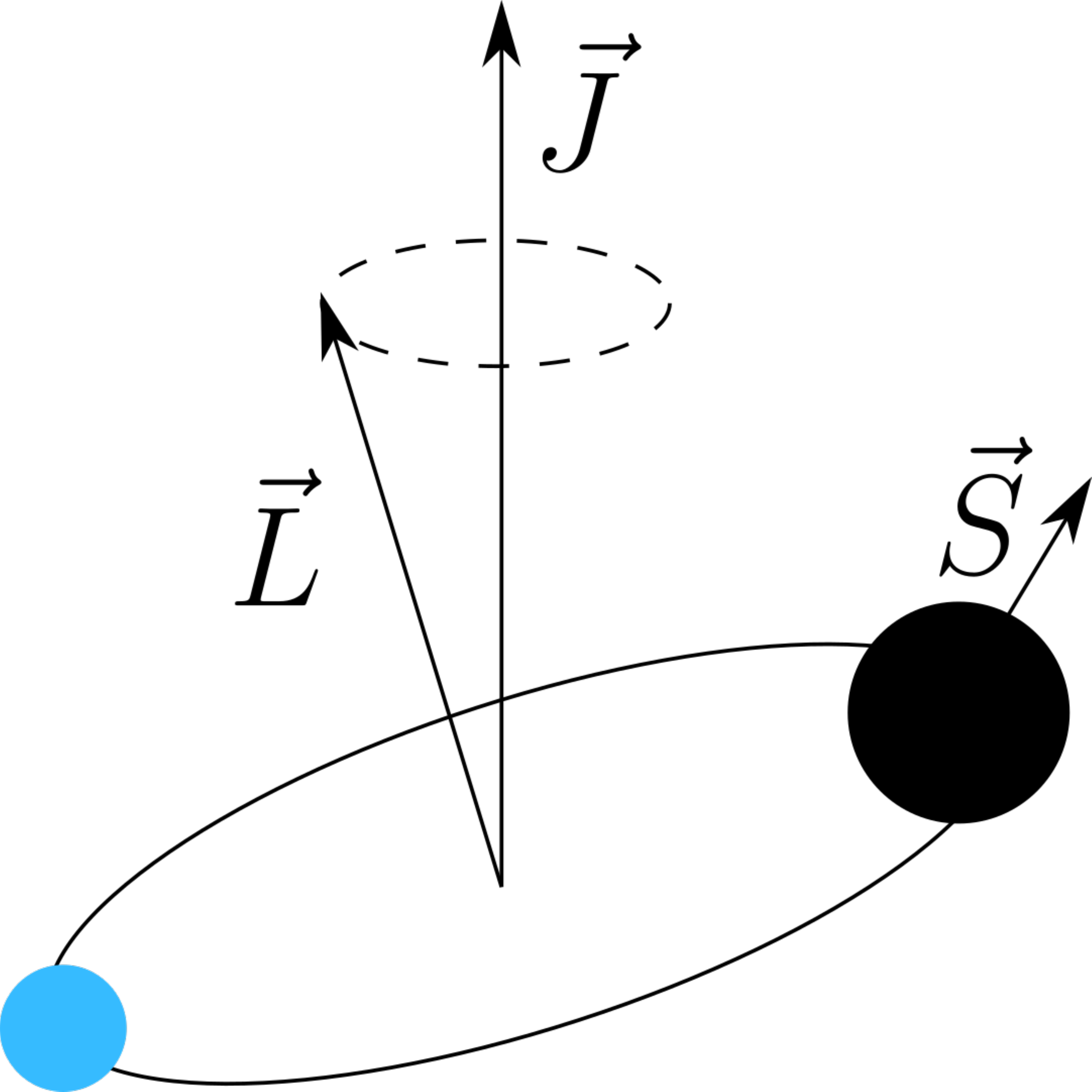}{Precessing binary.  $\vec{J}, \vec{L}$ and $\vec{S}$ are respectively total, orbital and spin angular momentum.}{FIG:concept_precession}
Since, for the case, the orbital plane precesses, the amplitude of the emitted GW is modulated~\cite{precession_KippThorne, precession1, precession2}.
By considering the modulation, the estimation precision of binary parameters should be improved more.

The organization of this paper is as follows.
In \secref{SEC:waveform}, waveforms with precession and the Doppler phases are reviewed.
In \secref{SEC:analysis}, a Fisher analysis for the parameter estimation of early warning is reviewed.
In \secref{SEC:result}, parameter estimation errors as a function of frequency or time to merger are shown.
\secref{SEC:conclusion} is devoted to a summary.

\section{Waveform} \label{SEC:waveform}
\begin{figure}[tbp]
	\begin{tabular}{c}
	\begin{minipage}{0.8\linewidth}
		\centering
		\includegraphics[width=\linewidth]{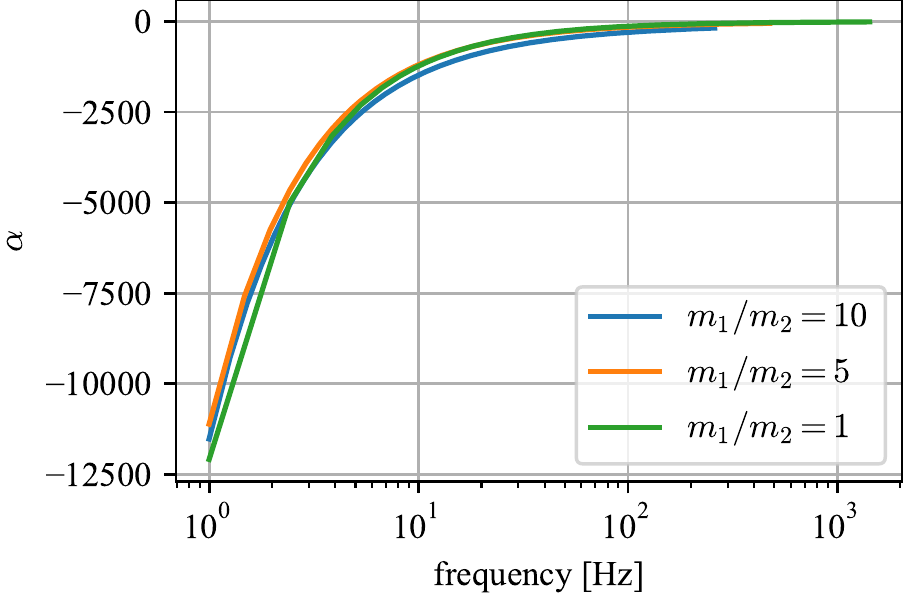} \\
		\vspace{3mm}
        \includegraphics[width=0.95\linewidth]{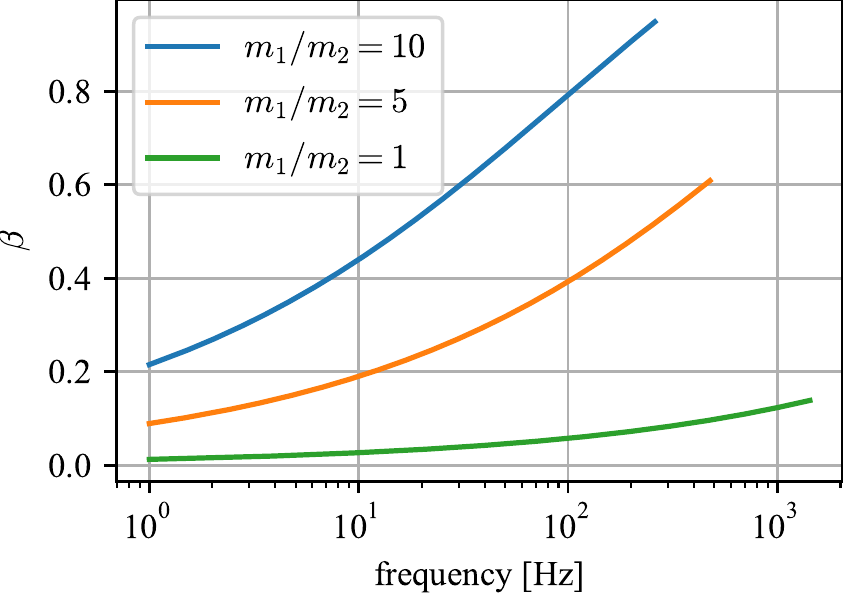} \\
		\vspace{3mm}
		\includegraphics[width=\linewidth]{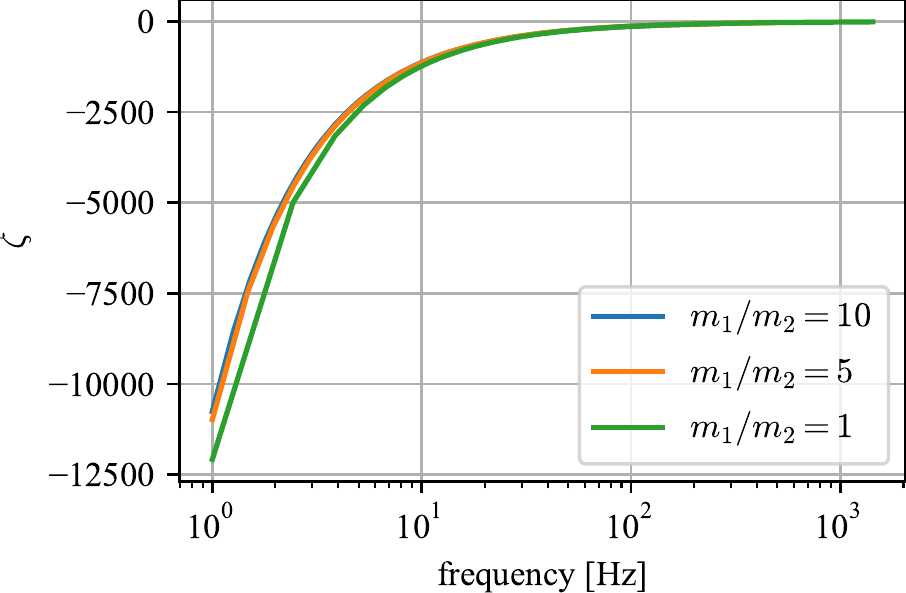} 
	\end{minipage}
	\end{tabular}
	\caption{Precession angles for different mass ratios as a function of frequency.  The top pannel is for $\alpha$, the middle pannel is for $\beta$, and the bottom pannel is for $\zeta$.}
	\label{FIG:precession_angles}
\end{figure}

Typically, a spin of a BH is much larger than that of a NS, because a spin angular momentum is proportional to mass and a BH is much heavier than a NS.
Thus, even if both the components in a binary have finite spins, the spins of NSs can be neglected.
To see the precession effect, a case in which a lighter component has no spin but a heavier component has non-zero spin is considered in this paper for simplicity.
Variations of total and spin angular momenta, $\vec{J}$ and $\vec{S}$, are at higher post-Newtonian (PN) order than the one of an orbital angular momentum, $\vec{L}$~\cite{precession_KippThorne}.
Then $\vec{L}$ is precessing around $\vec{J}$ (see \figref{FIG:concept_precession}).
Considering the coordinates whose $z$-direction is along $\vec{J}$, one can deal with the precession by a rotation matrix with the Euler angles, $\alpha$ and $\beta$ which express $\vec{L}$ relative to $\vec{J}$, and $\zeta$ which expresses how much angle the component objects are rotated around $\vec{L}$~\cite{precession1, precession2}.
They are
\eq{
	\alpha(v) &= \frac{\eta}{M} \left( 2 + \frac{3m_2}{2m_1} \right) \int^{v} v^5 \Gamma_J \left( \frac{\diff v}{\diff t} \right)^{-1} \,\diff v \, ,\\
	\beta(v) &= \cos^{-1} \left[ \frac{1+\kappa\gamma}{\Gamma_J} \right] \,, \label{EQ:beta} \\
	\zeta(v) &= \frac{\eta}{M} \left( 2 + \frac{3m_2}{2m_1} \right) \int^{v} v^5 (1 + \kappa \gamma) \left( \frac{\diff v}{\diff t} \right)^{-1} \,\diff v \,,
}
where
\eq{
	\gamma(v) &= \frac{|\vec{S}|}{|\vec{L}|} = \frac{m_1 \chi}{m_2} v \, , \label{EQ:gamma}\\
	\Gamma_J(v) &= \frac{|\vec{J}|}{|\vec{L}|} = \sqrt{1+2\kappa\gamma+\gamma^2} \, ,\\
	\kappa &= \frac{\vec{L}}{|\vec{L}|} \cdot \frac{\vec{S}}{|\vec{S}|} \, ,
}
$m_1$ and $m_2$ are heavier and lighter component masses, $\eta$ is the symmetric mass ratio, $M = m_1 + m_2$ is the total mass, $\chi$ is the dimensionless spin of the primary component, and 
$\diff v / \diff t$ up to $1.5${PN} is \cite{dvdt1, dvdt2}
\eq{
    \frac{\diff v}{\diff t} =& \frac{32}{5} \frac{\eta}{M} v^9 \left[ 1 - \left( \frac{743}{336} + \frac{11}{4} \eta \right) v^2 + 4\pi v^3 \right] \;.
}
For some mass ratio, the precession angles, $\alpha$, $\zeta$, and $\beta$, are shown in \figref{FIG:precession_angles}

By using these parameters, a GW signal in the Fourier domain is
\eq{
	\tilde{h}_I(f) &= Z_I(f) A(f)\, \e^{2\iu (\Phi_\mathrm{orb} - \zeta + \phi_{I, \mathrm{Doppler}})} \, , \label{EQ:waveform} \\
	A(f) &= \left( \frac{5\pi}{24} \right)^{1/2} \frac{G^2 \mathcal{M}^2}{c^5 d_L} \left( \frac{\pi G\mathcal{M} f}{c^3} \right)^{-7/6} \, ,\\
	Z_I(f) &= (C_+F_{I, +} + C_\cross F_{I, \cross}) - \iu(S_+F_{I, +} + S_\cross F_{I, \cross}) \label{EQ:z}
}
where $\mathcal{M}$ is the chirp mass, $d_L$ is the luminosity distance, $t_c$ and $\phi_c$ are the coalescence time and phase, $\phi_{I, \mathrm{Doppler}}$ is the Doppler phase between the $I$-th detector and the geocenter, $\Phi_\mathrm{orb}$ is the orbital phase from~\cite{PNexample1, PNexample2}.
To recognize $\kappa$ as constant, we neglect higher terms than $1.5$PN in $\Phi_\mathrm{orb}$ and use
\begin{widetext}
\eq{
    \Phi_\mathrm{orb} =& 2\pi ft_c - \phi_c - \pi /4 + \frac{3}{128 \eta} (\pi Mf)^{-5/3} \left[ 1 + \left( \frac{3715}{756} + \frac{55}{9} \eta \right) (\pi Mf)^{2/3} \right. \nonumber \\
    &\left. + \left\{ -16\pi + \left( \frac{113}{3} \left(1+\sqrt{1 - 4\eta} \right) - \frac{76}{3} \eta \right) \frac{\chi}{2} \right\} (\pi Mf) \right] \;.
}
The amplitude modulation factor for $I$-th detector, $Z_I(f)$, includes the antenna response functions for $+/\cross$-modes, $F_{I, +/\cross}$, and mixes them with the coefficients \cite{precession2}:
\eq{
	C_+ &= \frac{1}{2} \cos\theta_J [\sin^2\alpha - \cos^2\beta\cos^2\alpha] + \frac{1}{2} (\cos^2\beta\sin^2\alpha - \cos^2\alpha) - \frac{1}{2} \sin^2\theta_J\sin^2\beta - \frac{1}{4} \sin2\theta_J\sin2\beta\cos\alpha \, ,\\
	S_+ &= \frac{1}{2} (1 + \cos^2\theta_J)\cos\beta\sin2\alpha + \frac{1}{2}\sin2\theta_J\sin\beta\sin\alpha \, ,\\
	C_\cross &= -\frac{1}{2} \cos\theta_J(1 + \cos^2\beta)\sin2\alpha - \frac{1}{2} \sin\theta_J\sin2\beta\sin\alpha \, ,\\
	S_\cross &= -\cos\theta_J\cos\beta\cos2\alpha - \sin\theta_J\sin\beta\cos\alpha \,,
}
\end{widetext}
where $\theta_J$ is the inclination angle of $\vec{J}$. $v$ is related to $f$ by the Kepler's third law: $v(f) = \left( \pi GMf \right)^{1/3}$.
There are many independent parameters in the GW waveform:
\eq{
	\{ \lambda_i \} = \{ \mathcal{M}, \eta, \chi, t_c, \phi_c, d_L, \theta, \phi, \psi, \theta_J, \kappa, \alpha_0 \} \;, \label{EQ:parameter_set} 
}
where $\theta$ and $\phi$ are the longitude and latitude of a GW source, $\psi$ is the polarization angle, and $\alpha_0$ is the initial precession angle.
The initial value of $\zeta$ is absorbed by $\phi_c$.

Also, since the time for which a GW is observed in the bandwidth is $\mathcal{O}(\SI{1}{day})$, the GW signal is modulated by the rotation of the Earth.
By rotating $\phi$ with $\omega_E = 2\pi / (\SI{24}{hours})$, one can deal with the modulation.
That is, all $\phi$ included in Eq.~\eqref{EQ:waveform} should be replaced with $\phi + \omega_E (t_c - t)$.

\section{Analysis} \label{SEC:analysis}
{SNR} is defined as follows:
\eq{
	\mathrm{SNR}^2(f) = 4 \int_{f_\mathrm{min}}^{f} \frac{|\tilde{h}(f')|^2}{S_n(f')} \,\diff f' \label{EQ:snr} \;,
}
where $f_\mathrm{min} = \SI{1}{Hz}$ for the 3G detectors, and $S_n(f)$ is the power spectral density (see \figref{FIG:PSD}).
Here, the maximum cutoff frequency is set to the innermost stable circular orbit (ISCO) frequency.

To compute the parameter errors, the Fisher analysis~\cite{fisher_CBC} is used in this paper.
The Fisher matrix is defined as ($i,j$ run over Eq.~\eqref{EQ:parameter_set})
\eq{
	F'_{ij}(f) &= \left( \left. \frac{\partial h}{\partial \lambda_i} \right| \frac{\partial h}{\partial \lambda_j} \right)_f \, ,\\
	(a | b)_f &= 4 \Re \int_{f_\mathrm{min}}^{f} \frac{\tilde{a}^*(f') \tilde{b}(f')}{S_n(f')} \,\diff f' \, .
}
The Fisher analysis outputs just parameter errors, which sometimes can be too large beyond the physical ranges of parameters, for example, $\varDelta\theta > 2\pi, \varDelta\chi > 1$.
This is especially for low frequencies at which information from a signal is limited, because some of the parameters are degenerated with each other and the parameter estimation does not work at all. To obtain the physically relevant errors, Gaussian priors are added as follows:
\eq{
	F_{ij}(f) &= \left( \left. \frac{\partial h}{\partial \lambda_i} \right| \frac{\partial h}{\partial \lambda_j} \right)_f + \frac{\delta_{ij}}{(\delta \lambda_i)^2} \;,
}
where $\delta_{ij}$ is the Kronecker delta, and $\delta \lambda_i$ is a 1-$\sigma$ error of a Gaussian distribution and is taken as a physical range of the parameter.
By these priors, the probabilities of obtaining unphysical results decay exponentially~\cite{fisher}.
The inverse Fisher matrix is the covariance matrix, $\bm{\Sigma} = \bm{F}^{-1}$.
Thus, the standard deviation around the true value $\varDelta \lambda_i$ is
\eq{
	\varDelta \lambda_i = \sqrt{\Sigma_{ii}} \,,
}
and the sky localization error $\varDelta\Omega$ and error volume $\varDelta V$ are given by
\eq{
	\varDelta\Omega &= 2\pi|\sin\theta| \sqrt{\begin{vmatrix} \Sigma_{\theta \theta} & \Sigma_{\theta \phi} \\ \Sigma_{\phi \theta} & \Sigma_{\phi \phi} \end{vmatrix}} \, ,\\
	\varDelta V &= \frac{4}{3} \pi |\sin\theta| \sqrt{\begin{vmatrix} \Sigma_{d_L d_L} & d_L \Sigma_{d_L \theta} & d_L \Sigma_{d_L \phi} \\ d_L \Sigma_{\theta d_L} & d_L^2 \Sigma_{\theta \theta} & d_L^2 \Sigma_{\theta \phi} \\ d_L \Sigma_{\phi d_L} & d_L^2 \Sigma_{\phi \theta} & d_L^2 \Sigma_{\phi \phi} \\ \end{vmatrix}} \, .
}

Following the time evolution of $d_L, \varDelta\Omega, \varDelta V$ with $t_c - t = \frac{5}{256} (\pi f)^{-8/3} (\frac{G \mathcal{M}}{c^3})^{-5/3}$ at the Newtonian order~\cite{maggiore1, Jolien}, the performance of early warning can be estimated.

In this paper, 200 samples are calculated for three cases of {NSBH} and {BNS} with following parameters: $z = 0.1$, $\chi = 0.5$, $\kappa = 0$, $m_2 = \SI{1.4}{M_\odot}$, and $m_1 = \SI{14}{M_\odot}, \SI{7}{M_\odot}$ and $\SI{1.41}{M_\odot}$ their mass ratios are $m_1/m_2=10$, $5$, and $1$).
The angles, $\theta$ and $\phi$, are sampled isotropically over the sky and $\psi$ and $\cos\theta_J$ are uniformly.
It is assumed that there are one ET at the Virgo site and one CE at the Hanford site.
To obtain results for other redshifts, $z \neq 0.1$, the SNR scalings in Eqs.~\eqref{EQ:scale_dL} -- \eqref{EQ:scale_V} can be used.

\section{Result} \label{SEC:result}
In this section, we show the results of parameter estimation for NSBH with the mass ratios, $m_1 / m_2 = 10$ and $5$, observed with a single detector (ET) and two detectors (ET \& CE). We also compare the results with those for BNS in order to check consistency with the previous works and see how much the results change from the case of NSBH.

\subsection{Time evolutions}
\begin{figure*}[tbhp]
	\begin{tabular}{cc}
	\begin{minipage}{0.4\linewidth}
		\centering
		\includegraphics[width=\linewidth]{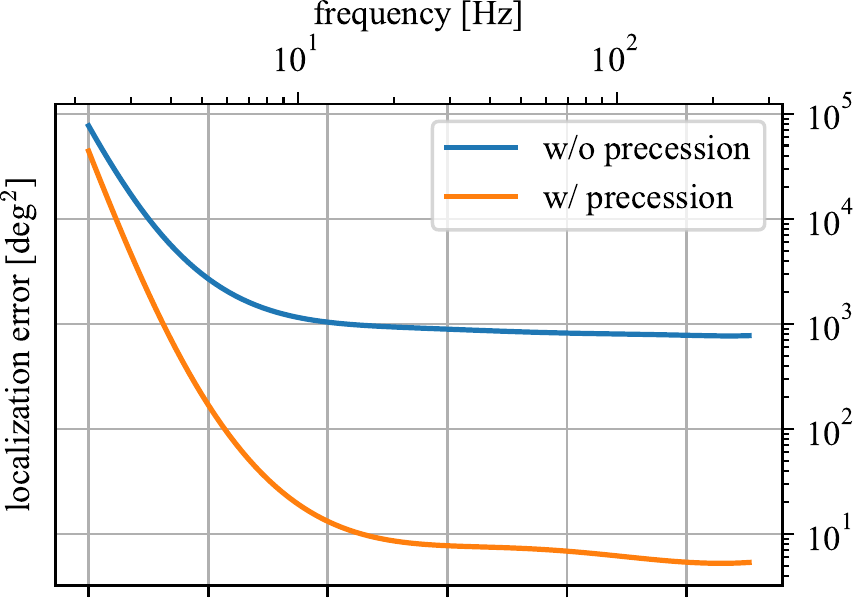} \\
		\includegraphics[width=\linewidth]{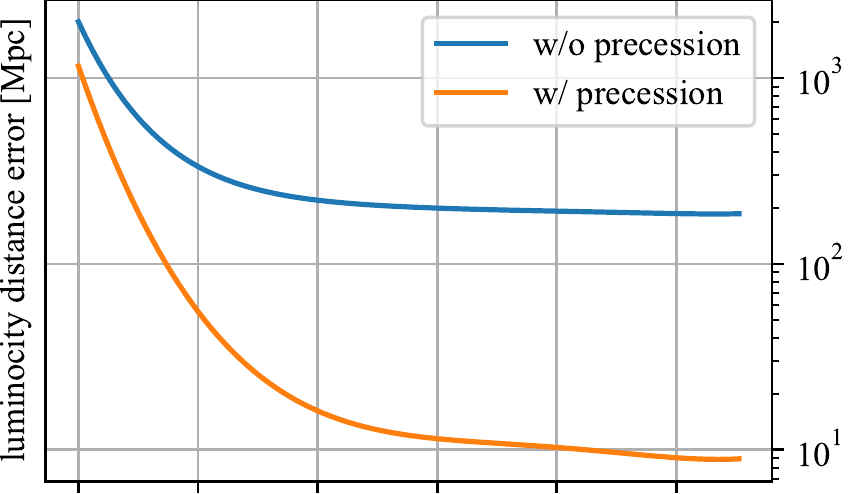} \\
		\includegraphics[width=\linewidth]{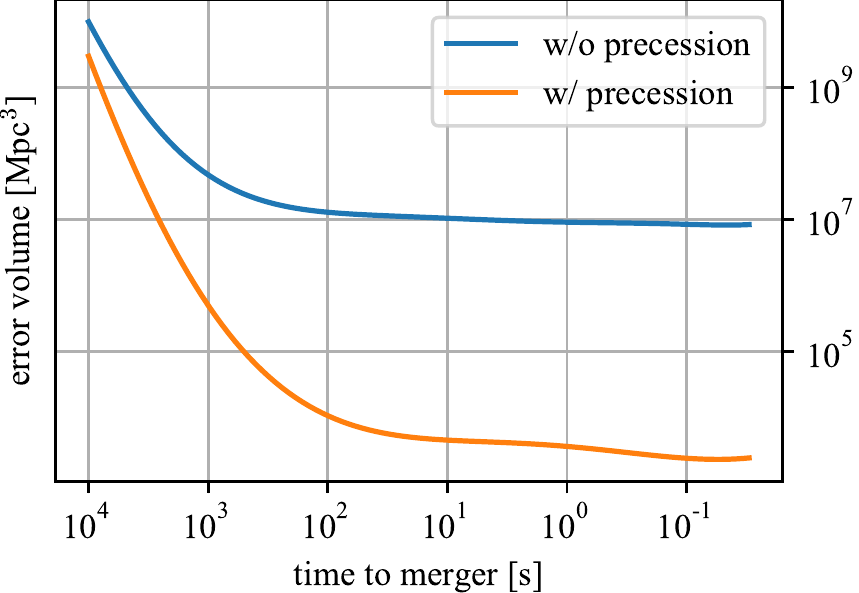}
	\end{minipage}
	\hspace{5mm}
	\begin{minipage}{0.4\linewidth}
		\centering
		\includegraphics[width=\linewidth]{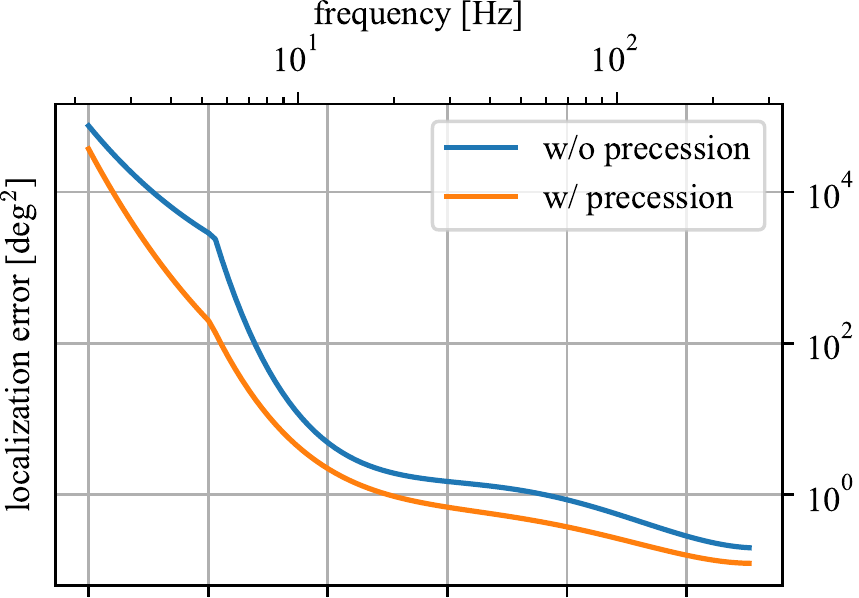} \\
		\includegraphics[width=\linewidth]{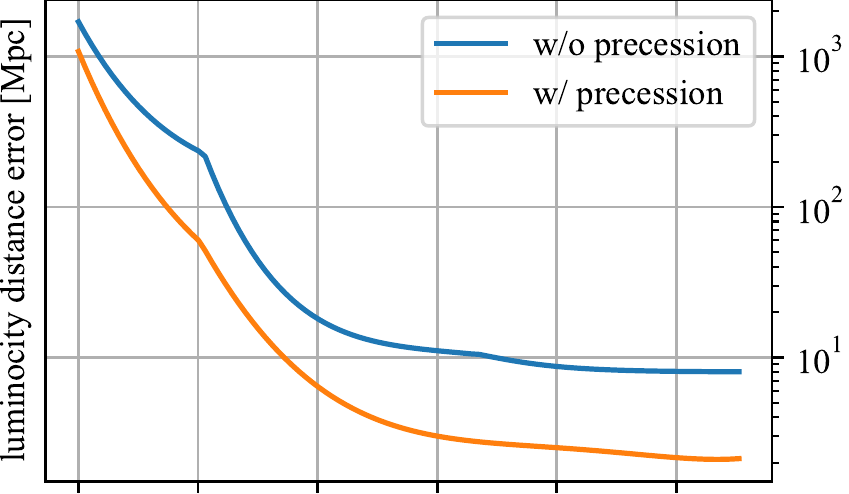} \\
		\includegraphics[width=\linewidth]{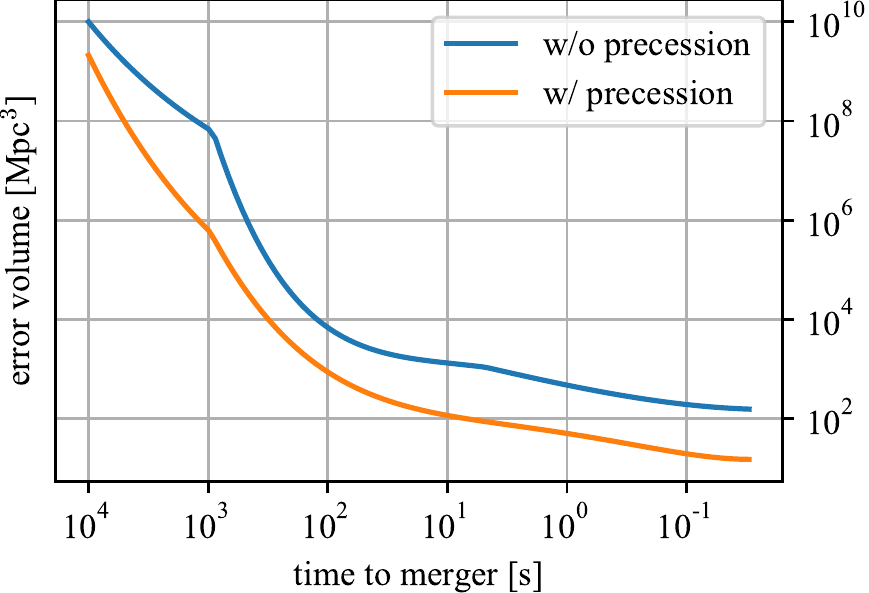}
	\end{minipage}
	\end{tabular}
	\caption{Medians of localization errors for NSBH with $m_1 / m_2 = 10$ as a function of time to merger.  Left column is for the single detector case (ET). Right column is for the double detector case (ET \& CE). From the top to the bottom, the sky localization error, the distance error, and the error volume as a function of time to merger.  Blue (Orange) lines are without (with) a precession effect.}
	\label{FIG:result_NSBH_TimeEvolution_q10}
\end{figure*}

\begin{figure*}[tbhp]
	\begin{tabular}{cc}
	\begin{minipage}{0.4\linewidth}
		\centering
		\includegraphics[width=\linewidth]{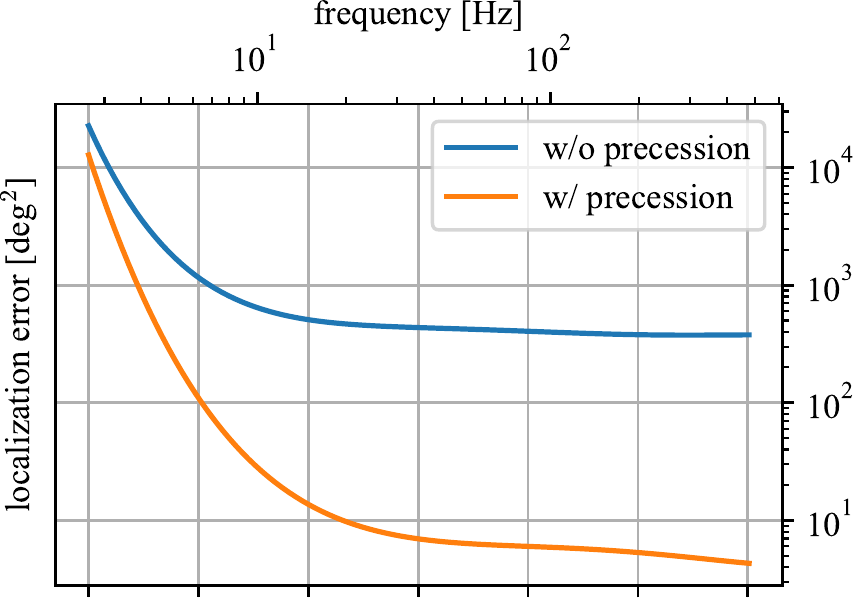} \\
		\includegraphics[width=\linewidth]{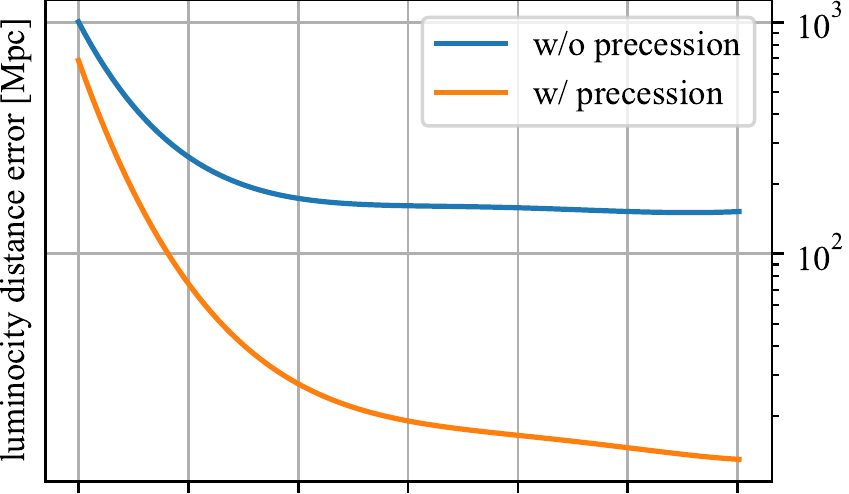} \\
		\includegraphics[width=\linewidth]{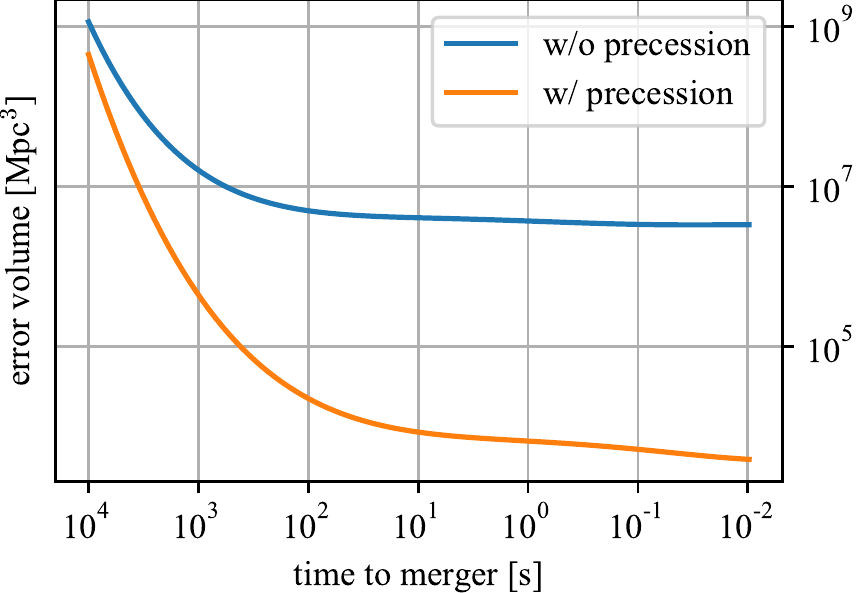}
	\end{minipage}
	\hspace{5mm}
	\begin{minipage}{0.4\linewidth}
		\centering
		\includegraphics[width=\linewidth]{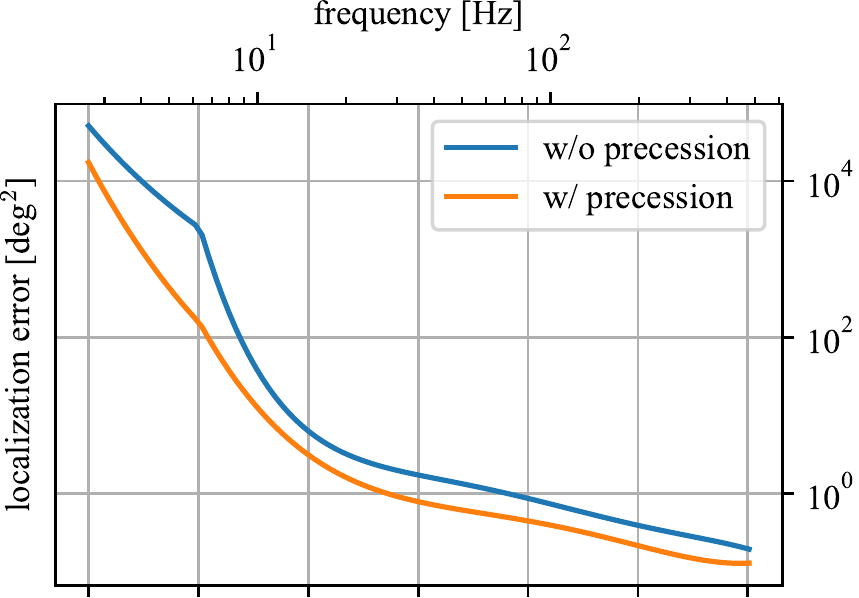} \\
		\includegraphics[width=\linewidth]{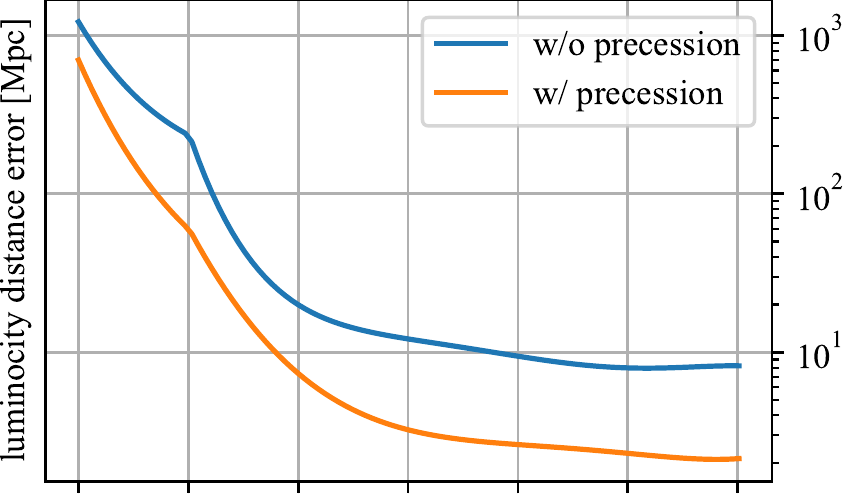} \\
		\includegraphics[width=\linewidth]{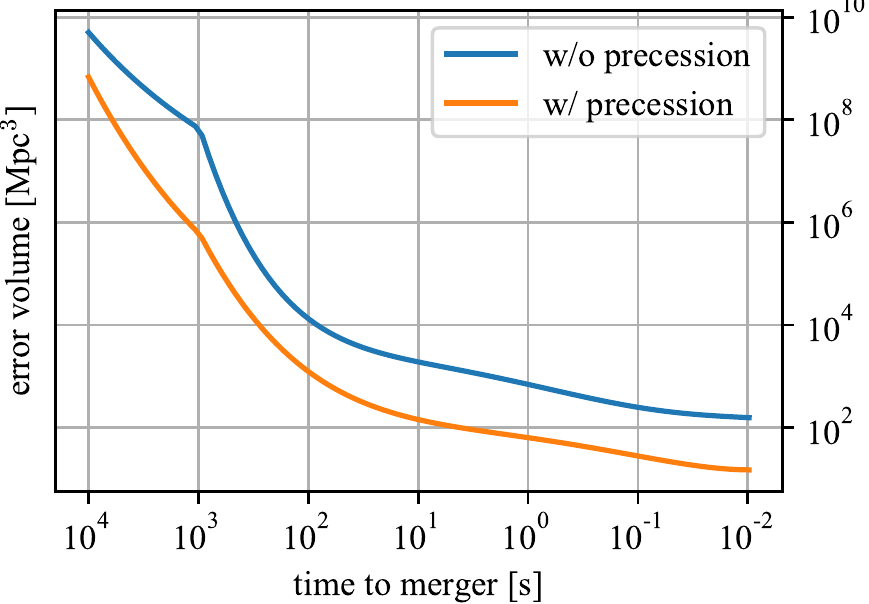}
	\end{minipage}
	\end{tabular}
	\caption{The same as Fig.~\ref{FIG:result_NSBH_TimeEvolution_q10} but for $m_1 / m_2 = 5$.}
	\label{FIG:result_NSBH_TimeEvolution_q5}
\end{figure*}

Figures~\ref{FIG:result_NSBH_TimeEvolution_q10} and \ref{FIG:result_NSBH_TimeEvolution_q5} are medians of the sky localization errors $\varDelta\Omega$, the distance errors $\varDelta d_L$, and the error volumes $\varDelta V$ for $m_1 / m_2 = 10$ and $5$, respectively.
In the single detector case (ET) of Figs.~\ref{FIG:result_NSBH_TimeEvolution_q10} and \ref{FIG:result_NSBH_TimeEvolution_q5}, the precession effect much improves the localization errors at a fixed time to merger (see \tabref{TAB:NSBH}).
Furthermore, final localization errors at the time of merger are also much improved, reaching $\SI{13}{Mpc}$, $\SI{5.7}{deg^2}$, and $2.5 \times 10^3 \unit{Mpc^3}$ for $m_1 / m_2 = 10$ and $\SI{13}{Mpc}$, $\SI{4.5}{deg^2}$, and $2.3 \times 10^3 \unit{Mpc^3}$ for $m_1 / m_2 = 5$, whose volumes include $25$ and $23$ galaxies on average, respectively, assuming the average number density of massive galaxies, $\SI{0.01}{Mpc^{-3}}$ as in, e.~g.,~ \cite{Measurement_HubbleConstant}.
The reason of the significant improvement for a single detector is breaking a degeneracy between the luminosity distance and the inclination angle of the orbital plane.
For a non-precessing case with a single detector, even if the overall amplitude of a GW is estimated, it is difficult to separate those parameters, because the inclination angle is constant in time.
In contrast, for a precessing case, the degeneracy is broken because the inclination angle is varying in time.

For double detector case (ET \& CE) in Figs.~\ref{FIG:result_NSBH_TimeEvolution_q10} and \ref{FIG:result_NSBH_TimeEvolution_q5}, the results are less improved than the single detector case.
However, the final errors with a precession effect reach $\SI{2.1}{Mpc}$, $\SI{0.12}{deg^2}$, and $\SI{13}{Mpc^3}$ for for $m_1 / m_2 = 10$ and $\SI{2.1}{Mpc}$, $\SI{0.13}{deg^2}$, and  $\SI{12}{Mpc^3}$ for for $m_1 / m_2 = 5$, whose volumes include $0.13$ and $0.12$ galaxies on average and enable us to identify a unique galaxy.
The reason of the less improvement for the double detector case is because the degeneracy is already partially broken by measuring a GW with differently orientating detectors.

\begin{table*}[tbhp]
	\caption{Time to merger for NSBH with some conditions from Figs.~\ref{FIG:result_NSBH_TimeEvolution_q10} and \ref{FIG:result_NSBH_TimeEvolution_q5}.  The number density of massive galaxies assumed is $\SI{0.01}{Mpc^{-3}}$.}
	\begin{tabular}{cl|cc|cc} \hline \hline
		detector & & \multicolumn{4}{c}{time to merger [s]} \\ \hline
		& & \multicolumn{2}{c}{$m_1 / m_2 = 10$} & \multicolumn{2}{c}{$m_1 / m_2 = 5$} \\ \hline
		& & w/o precession & w/ precession & w/o precession & w/ precession \\ \hline
		ET & $\varDelta\Omega = \SI{1000}{deg^2}$ & $71$ & $2.4 \times 10^3$ & $7.9 \times 10^2$ & $3.6 \times 10^3$ \\
		& $\varDelta\Omega = \SI{100}{deg^2}$ & - & $7.3 \times 10^2$ & - & $9.3 \times 10^2$ \\
		& $\varDelta\Omega = \SI{10}{deg^2}$ & - & $52$ & - & $49$ \\
		& $\varDelta V = 10^7 \unit{Mpc^3}$ & $6.2$ & $2.6 \times 10^3$ & $5.7 \times 10^2$ & $3.5 \times 10^3$ \\
		& ($10^5$ galaxy) & & & \\
		& $\varDelta V = 10^5 \unit{Mpc^3}$ & - & $5.1 \times 10^2$ & - & $4.2 \times 10^2$ \\
		& ($10^3$ galaxy) & & & \\
		& $\varDelta V = 10^4 \unit{Mpc^3}$ & - & $93$ & - & $20$ \\
		& ($10^2$ galaxy) & & & \\ \hline
		ET \& CE & $\varDelta\Omega = \SI{100}{deg^2}$ & $4.1 \times 10^2$ & $7.8 \times 10^2$ & $4.2 \times 10^2$ & $8.2 \times 10^2$ \\
		& $\varDelta\Omega = \SI{10}{deg^2}$ & $1.6 \times 10^2$ & $3.0 \times 10^2$ & $1.4 \times 10^2$ & $2.5 \times 10^2$ \\
		& $\varDelta\Omega = \SI{1}{deg^2}$ & $1.5$ & $31$ & $1.6$ & $20$ \\
		& $\varDelta V = 10^5 \unit{Mpc^3}$ & $2.8 \times 10^2$ & $6.3 \times 10^2$ & $2.4 \times 10^2$ & $6.1 \times 10^2$ \\
		& ($10^3$ galaxy) & & & \\
		& $\varDelta V = 10^4 \unit{Mpc^3}$ & $1.2 \times 10^2$ & $3.1 \times 10^2$ & $83$ & $2.8 \times 10^2$ \\
		& ($10^2$ galaxy) & & & \\
		& $\varDelta V = 10^3 \unit{Mpc^3}$ & $4.1$ & $1.1 \times 10^2$ & $2.2$ & $86$ \\
		& ($10^1$ galaxy) & & & \\ \hline
	\end{tabular}
	\label{TAB:NSBH}
\end{table*}

\begin{figure*}[tbhp]
	\begin{tabular}{cc}
	\begin{minipage}{0.4\linewidth}
		\centering
		\includegraphics[width=0.98\linewidth]{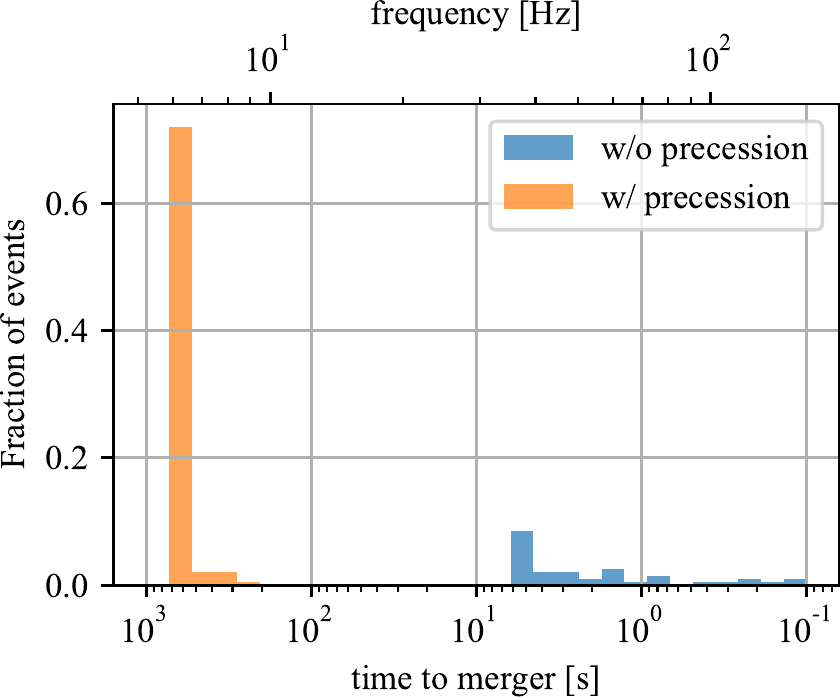}\\
		\vspace{3mm}
		\includegraphics[width=0.98\linewidth]{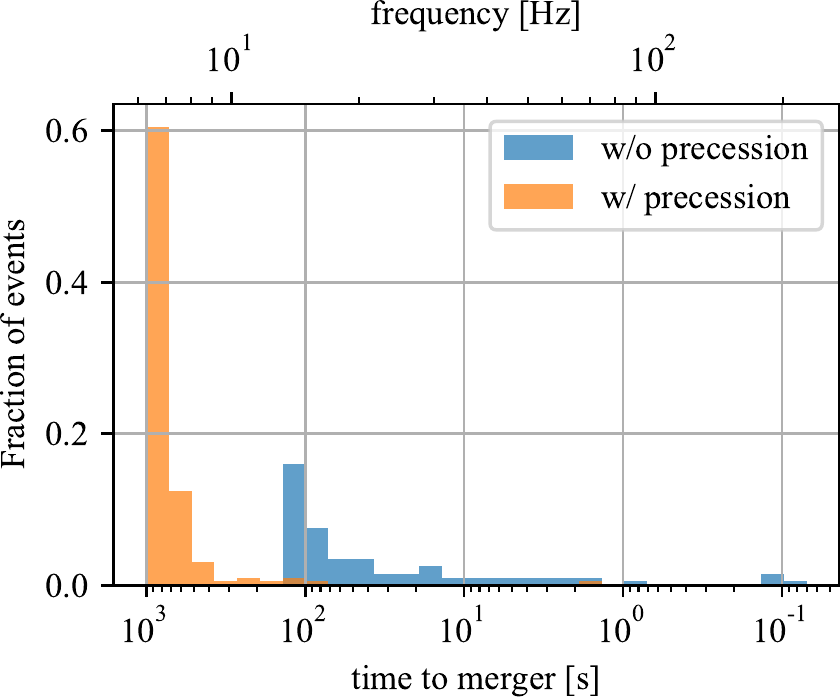}
	\end{minipage}
	\hspace{5mm}
	\begin{minipage}{0.4\linewidth}
		\centering
		\includegraphics[width=\linewidth]{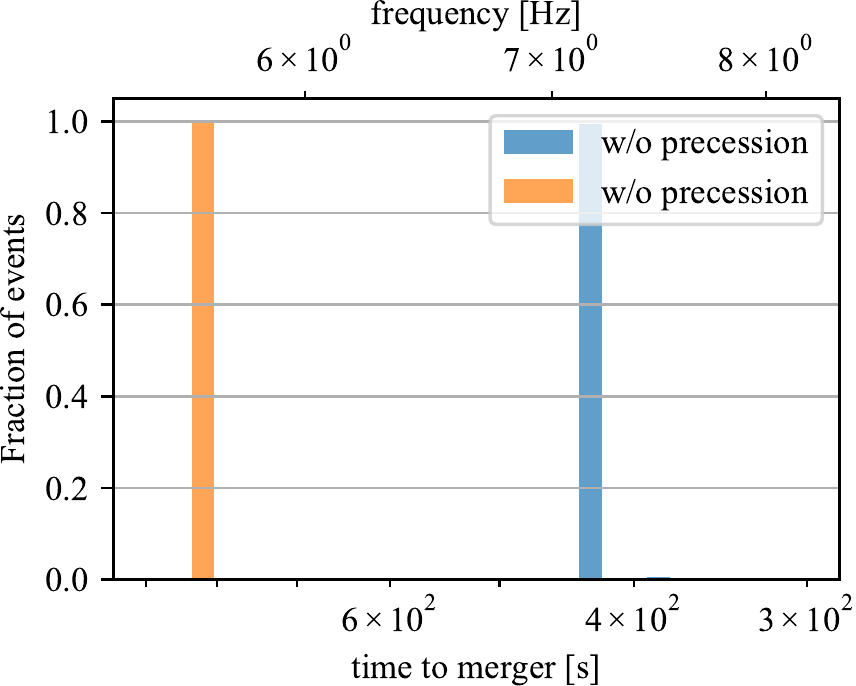}\\
		\vspace{3mm}
		\includegraphics[width=\linewidth]{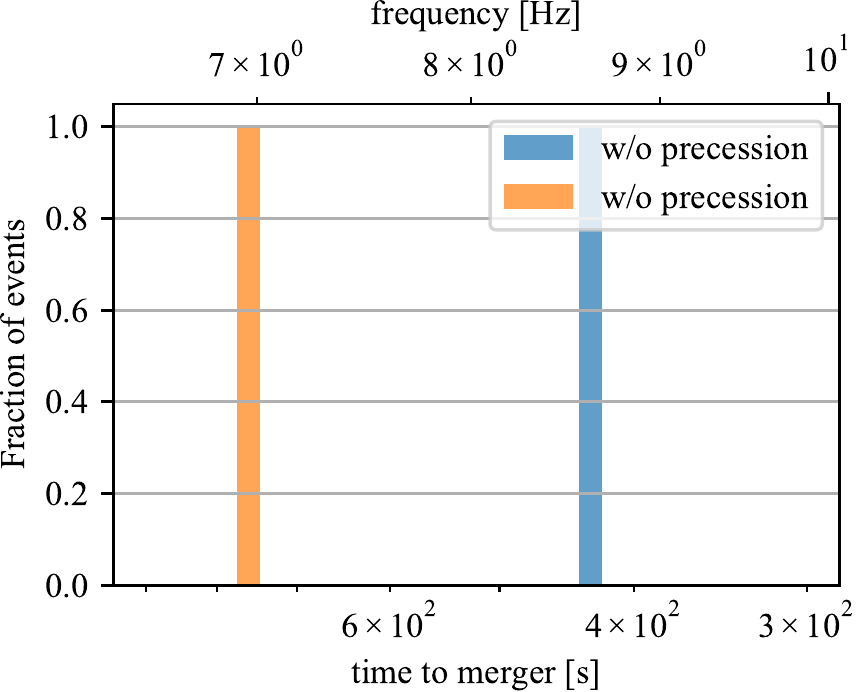}
	\end{minipage}
	\end{tabular}
	\caption{Distributions of time to merger of detectable events when sky localization reaches $\SI{100}{deg^2}$ for NSBH with $m_1 / m_2 = 10$ (top) and $5$ (bottom). Left column is for the single detector case (ET). Right one is for the double detector case (ET \& CE).}
	\label{FIG:result_NSBH_histogram_fraction_q10}
\end{figure*}

Variations of the samples when sky localization reaches $\SI{100}{deg^2}$ are shown in Fig.~\ref{FIG:result_NSBH_histogram_fraction_q10}.
One can understand that most of events with a precession effect are detected earlier than the ones without precession: for both mass ratios, by $\sim \SI{14}{minutes}$ for the single detector case and by $\sim \SI{5}{minutes}$ for the double detector case at best.
The variations of time to merger for the double detector case is smaller than that for the single detector case.
This is also because the degeneracy is broken.
The distributions of final errors are shown in \appref{SEC:fractions} for readers who are interested in how widely they are scattered.

One might wonder the bends of the curves around $\sim 10^3\unit{s}$ before merger for the double detector case. This behavior arises from the shapes of PSDs of ET and CE: ET is much more sensitive at low frequency region ($\lesssim \SI{10}{Hz}$) than CE. Then, for the low frequency range below $\sim 5\unit{Hz}$, it is effectively the same as the single detector case even there are two detectors, ET and CE.
Therefore, the precession effect is important for early warning, even if there are two 3G detectors (one is CE).

\subsection{Comparison with previous studies} \label{SEC:result_comparison}
As explained in \secref{SEC:waveform}, the spin of a NS can be neglected for high mass ratio systems.
However, there is no reason to be able to do same for a BNS case.
Also, for a BNS case, mass ratio is nearly unity, $\gamma(v)$ does not grow larger than a NSBH case, and then $\beta(v)$ does not, too (see Eqs.~\eqref{EQ:beta} and \eqref{EQ:gamma}).
That is, a precession effect is weaker for BNS.
Despite of the weakness, localization errors for the single detector case of ET are calculated to compare with the previous study (without a precession effect)~\cite{earlywarning_BNS} and to see how much the results change from the case of NSBH.

\begin{figure}[tbhp]
	\begin{tabular}{c}
	\begin{minipage}{0.8\linewidth}
		\centering
		\includegraphics[width=\linewidth]{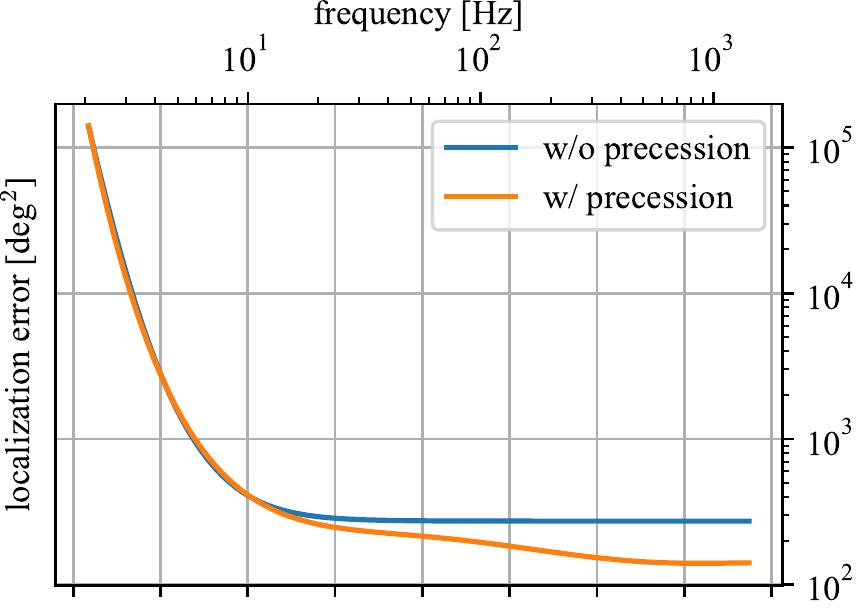} \\
		\includegraphics[width=\linewidth]{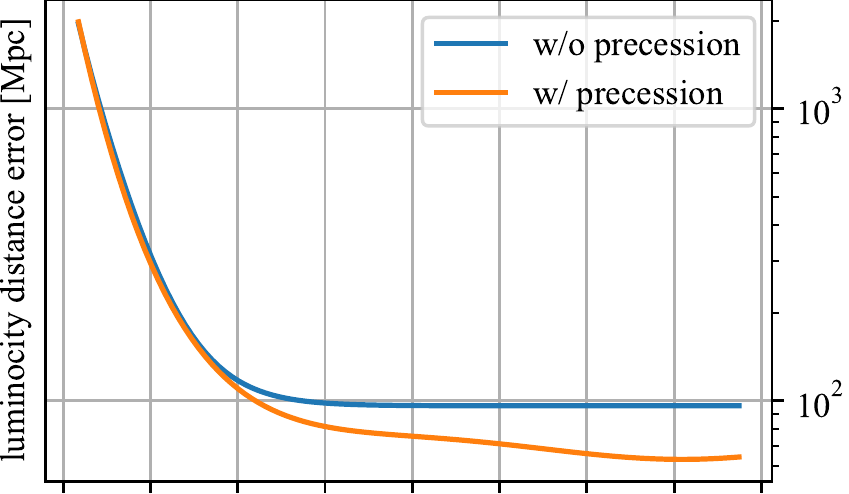} \\
		\includegraphics[width=\linewidth]{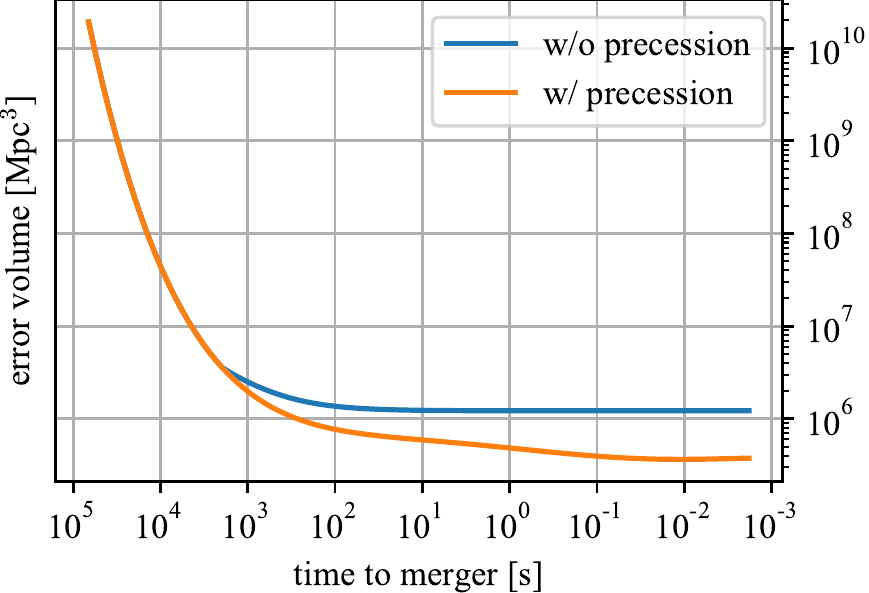}
	\end{minipage}
	\end{tabular}
	\caption{Medians of localization errors for BNS in the single detector case (ET) as a function of time to merger. From the top to the bottom, the sky localization error, the distance error, and the error volume as a function of time to merger. Blue (Orange) lines are without (with) the precession effect.}
	\label{FIG:result_BNS_TimeEvolution}
\end{figure}

Figure~\ref{FIG:result_BNS_TimeEvolution} is the results of the sky localization error $\varDelta\Omega$, the distance error $\varDelta d_L$, and the error volume $\varDelta V$ as a function of time to merger for the single detector case (ET).
The precession effect less improves the localization errors (see \tabref{TAB:BNS}) and their final values due to its weakness.
Especially, the final localization errors with the precession effect are $\SI{64}{Mpc}$ in distance, $1.4 \times 10^2 \unit{deg^2}$ in sky localization, $3.0 \times 10^5 \unit{Mpc^3}$ in volume, which includes $3.0 \times 10^3$ galaxies.
Although there are a few differences between this study and \cite{earlywarning_BNS} (SNR cutoff is $12$ instead of $8$ for this study, $d_L = \SI{400}{Mpc}$ instead of $z=0.1$ is assumed, $500$ sources instead of $200$ sources, and spinless instead of non-zero precessing spin), the result is consistent with TABLE~I in~\cite{earlywarning_BNS} within a factor of $\mathcal{O}(1)$, whose main origin is the spin values.

\begin{table}[tbhp]
	\caption{Time to merger for BNS with some conditions from \figref{FIG:result_BNS_TimeEvolution}.  The Earth rotation is considered.  A density of galaxies is $\SI{0.01}{Mpc^{-3}}$~\cite{Measurement_HubbleConstant}.}
	\begin{tabular}{cl|cc} \hline \hline
		detector & & \multicolumn{2}{c}{time to merger [s]} \\ \hline
		& & w/o precession & w/ precession \\ \hline
		ET & $\varDelta\Omega = \SI{1000}{deg^2}$ & $4.2 \times 10^3$ & $4.2 \times 10^3$ \\
		& $\varDelta\Omega = \SI{700}{deg^2}$ & $2.8 \times 10^3$ & $2.8 \times 10^3$ \\
		& $\varDelta\Omega = \SI{400}{deg^2}$ & $9.4 \times 10^2$ & $9.4 \times 10^2$ \\
		& $\varDelta V = 10^7 \unit{Mpc^3}$ & $4.4 \times 10^3$ & $4.4 \times 10^3$ \\
		& ($10^5$ galaxy) & \\
		& $\varDelta V = 6 \times 10^6 \unit{Mpc^3}$ & $3.1 \times 10^3$ & $3.1 \times 10^3$ \\
		& ($6 \times 10^4$ galaxy) & \\
		& $\varDelta V = 2 \times 10^6 \unit{Mpc^3}$ & $5.8 \times 10^2$ & $1.0 \times 10^3$ \\
		& ($6 \times 10^4$ galaxy) & \\ \hline
	\end{tabular}
	\label{TAB:BNS}
\end{table}

\section{Conclusion} \label{SEC:conclusion}
The constructions of 3G detectors are planned, especially ET and CE.
They are sensitive to GWs from $\sim \SI{1}{Hz}$ or $\sim \SI{1}{day}$ before mergers.
In this case, one can consider the modulation of a GW signal by the Earth rotation and a precession effect for NSBH.
In this paper, for the first time we estimate how much the performance of early warning is improved by considering the precession effect together with the Earth rotation.

By considering precession, since the degeneracy between the inclination angle and the luminosity distance is broken, the performance of early warning is much improved for the single detector case (see \tabref{TAB:NSBH}, \figref{FIG:result_NSBH_TimeEvolution_q10} and \figref{FIG:result_NSBH_TimeEvolution_q5}).
For the double detector case, because the degeneracy is already partially broken by measuring a GW with differently orientating detectors, the performance of early warning is less improved from the case without the precession effect (see \tabref{TAB:NSBH}, \figref{FIG:result_NSBH_TimeEvolution_q10} and \figref{FIG:result_NSBH_TimeEvolution_q5}).
Despite of the double detector case, it is effectively the single detector case for low frequencies below $\sim 5\unit{Hz}$ due to the difference of the power spectral densities between ET and CE (see \figref{FIG:PSD}).
This is the reason of inhomogeneous improvement of the performance in each frequency range for the double detector case.

In conclusion, the precession effect is very useful for the early warning, especially, when only one detector is in an observing mode.
Even if two detectors are in the observing modes, the precession effect modestly improves the performance of early warning.
Importantly, the precision of the sky localization can reach $100\unit{deg}^2$ and $\SI{10}{deg^2}$ at the time of $12$ -- $15 \unit{minutes}$ and $50$ -- $\SI{300}{seconds}$ before merger, respectively.
These small localization areas enable us to search an EM counterpart within the fields of view of EM telescopes, for example, Cherenkov telescope array (CTA)~\cite{sphradiometer, CTA_FOV} ($\sim \SI{70}{deg^2}$), Swift-BAT~\cite{Swift_BAT} ($\sim \SI{4600}{deg^2}$), Fermi-LAT~\cite{Fermi_LAT} ($\sim \SI{7900}{deg^2}$), and LSST~\cite{LSST} ($\sim \SI{9.6}{deg^2}$).

\begin{acknowledgments}
T.~T. is supported by International Graduate Program for Excellence in Earth-Space Science (IGPEES).
A.~N. is supported by JSPS KAKENHI Grant Nos. JP19H01894 and JP20H04726 and by Research Grants from Inamori Foundation.
S.~M. is supported by JSPS KAKENHI Grant Number 19J13840.
\end{acknowledgments}

\appendix
\section{Final Error Distributions} \label{SEC:fractions}
\begin{figure*}[tbhp]
	\begin{tabular}{cc}
	\begin{minipage}{0.4\linewidth}
		\centering
		\includegraphics[width=\linewidth]{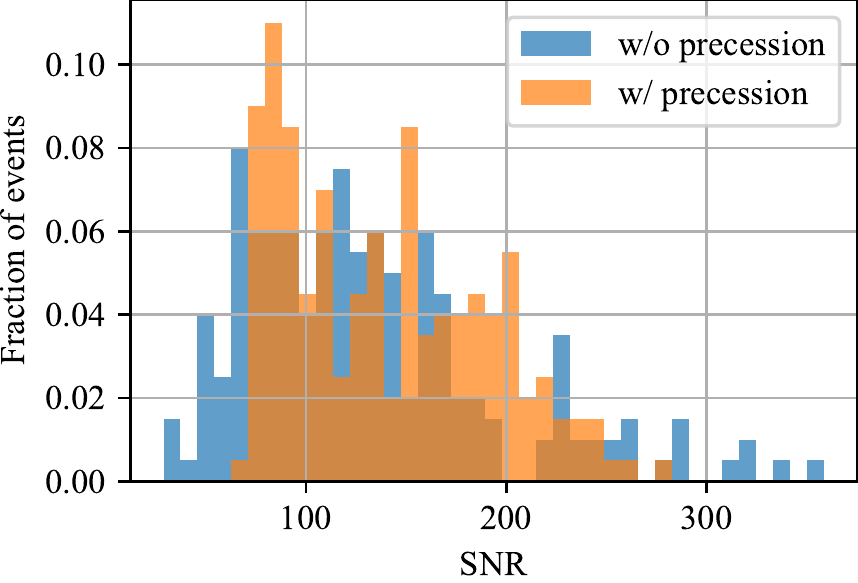}\\
		\vspace{2mm}
		\includegraphics[width=\linewidth]{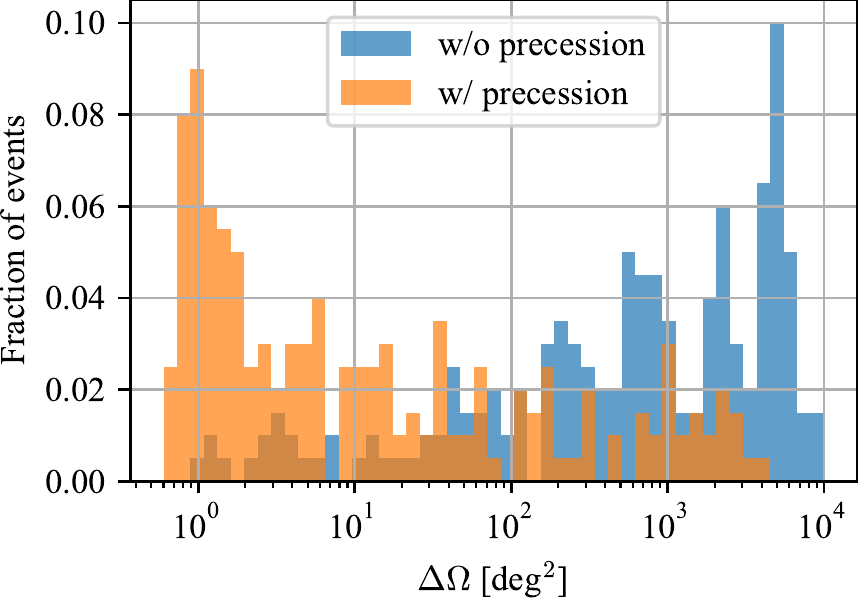}\\
		\vspace{2mm}
		\includegraphics[width=\linewidth]{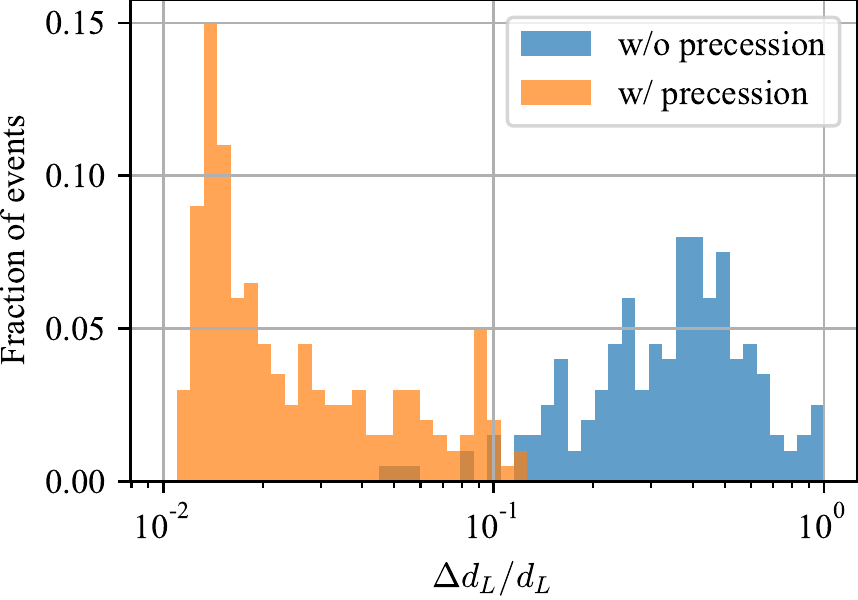}
	\end{minipage}
	\hspace{5mm}
	\begin{minipage}{0.4\linewidth}
		\centering
		\includegraphics[width=\linewidth]{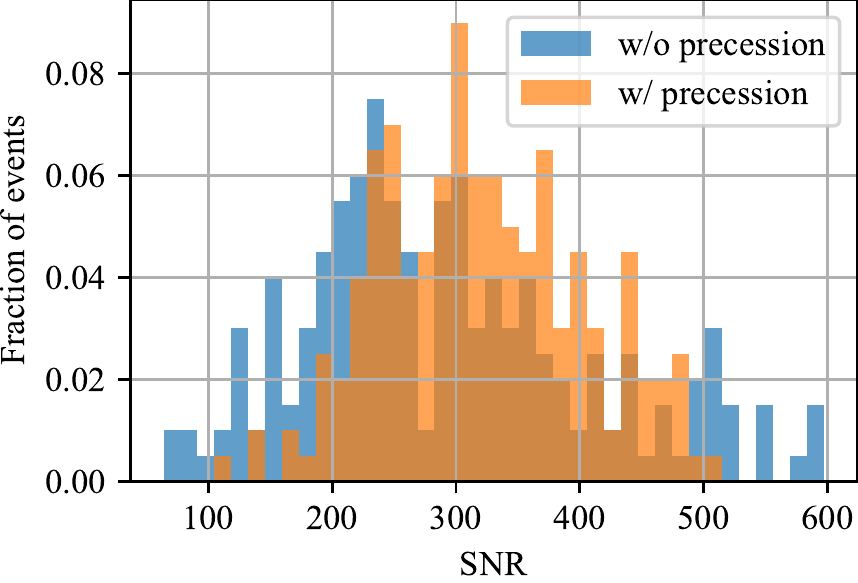}\\
		\vspace{2mm}
		\includegraphics[width=\linewidth]{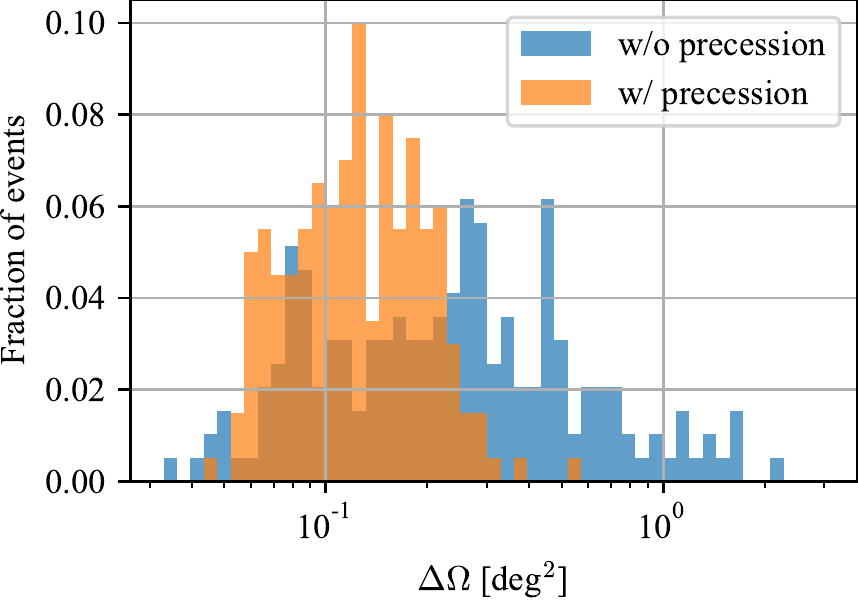}\\
		\vspace{2mm}
		\includegraphics[width=\linewidth]{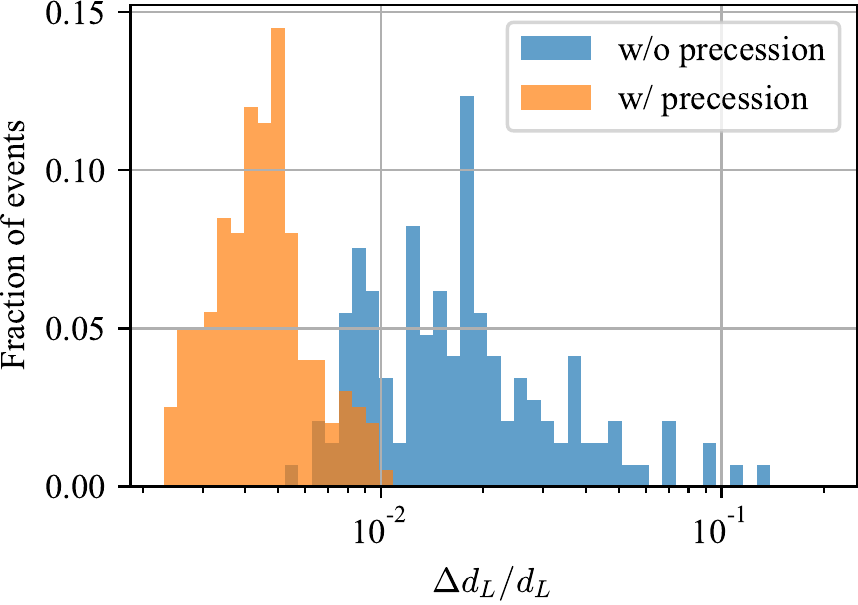}
	\end{minipage}
	\end{tabular}
	\caption{Distributions of SNR (top), $\varDelta\Omega$ (middle) and $\varDelta d_L / d_L$ (bottom) for NSBH with $m_1 / m_2 = 10$. Left column is for the single detector case (ET). Right one is for the double detector case (ET \& CE).}
	\label{FIG:fraction_NSBH_snr}
\end{figure*}

\begin{figure*}[tbhp]
	\begin{tabular}{ccc}
		\centering
		\includegraphics[width=0.3\linewidth]{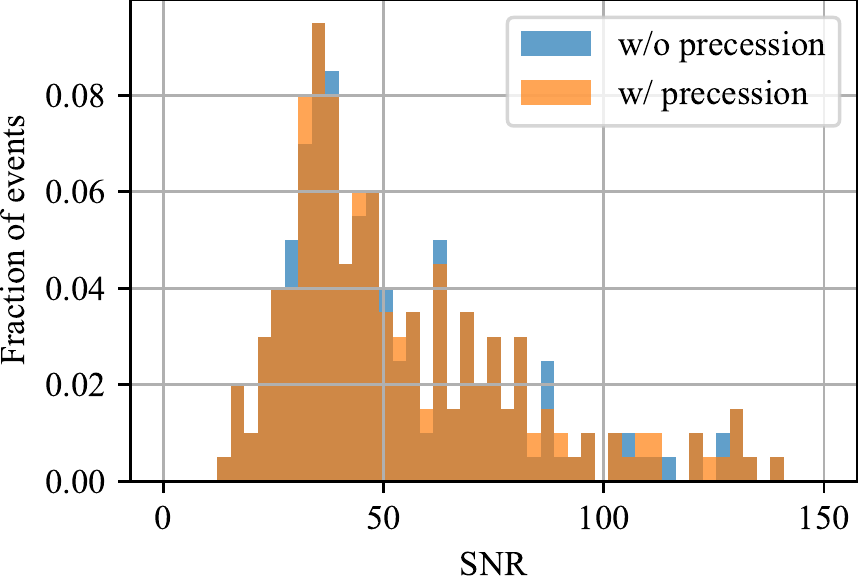}
		\hspace{2mm}
		\includegraphics[width=0.3\linewidth]{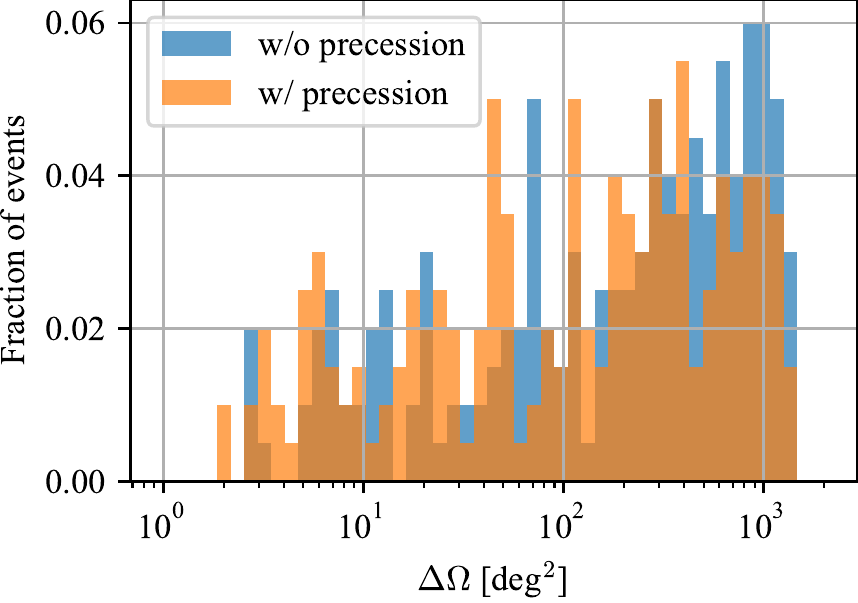}
		\hspace{2mm}
		\includegraphics[width=0.3\linewidth]{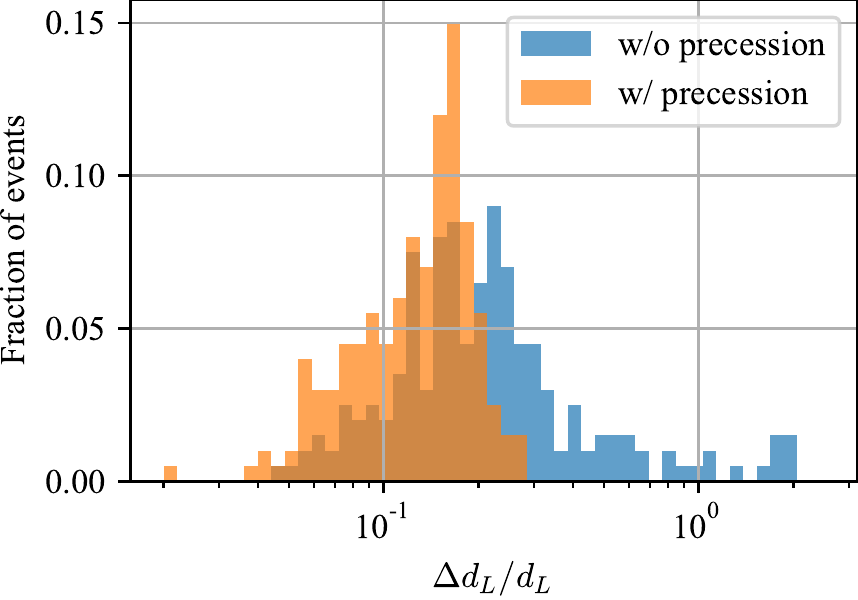}
	\end{tabular}
	\caption{Distributions of SNR (left), $\varDelta\Omega$ (middle) and $\varDelta d_L / d_L$ (right) for BNS.}
	\label{FIG:fraction_BNS_snr}
\end{figure*}

Figure~\ref{FIG:fraction_NSBH_snr} is the distributions of SNR and the localization errors at the merger (up to the ISCO frequency) for NSBH with $m_1 / m_2 = 10$ samples calculated in \secref{SEC:result}.
The left panels are for the single detector case, and the right panels are for the double detectors case.
Figure~\ref{FIG:fraction_BNS_snr} is the distributions for BNS.
Blue histograms are for non-precessing case, and orange histograms are for a precessing case.

When binaries are precessing, the inclination angle does not stay in a certain direction that GW amplitude becomes larger or smaller.
Thus the variation of SNR for NSBH becomes smaller when precession is considered, because the precession effect is stronger for NSBH than for BNS.
The localization errors, $\varDelta\Omega$ and $\varDelta d_L / d_L$, for NSBH and BNS are improved by including the precession effect.

\section{SNR scalings} \label{SEC:snr_scaling}
Errors estimated by the Fisher analysis scale with respect to SNR~\cite{fisher_CBC}:
\eq{
	\varDelta d_L &\propto \mathrm{SNR}^{-1} \, , \label{EQ:scale_dL}\\
	\varDelta\Omega &\propto \mathrm{SNR}^{-2} \, , \label{EQ:scale_Omega}\\
	\varDelta V &\propto \mathrm{SNR}^{-3} \, . \label{EQ:scale_V}
}
For sources at redshifts other than $z=0.1$, all the errors estimated in this paper basically scale by the relations above, though the range of the frequency integral is slightly modified because of the redshift.

\begin{figure*}[tbhp]
	\begin{tabular}{cc}
	\begin{minipage}{0.4\linewidth}
		\centering
		\includegraphics[width=\linewidth]{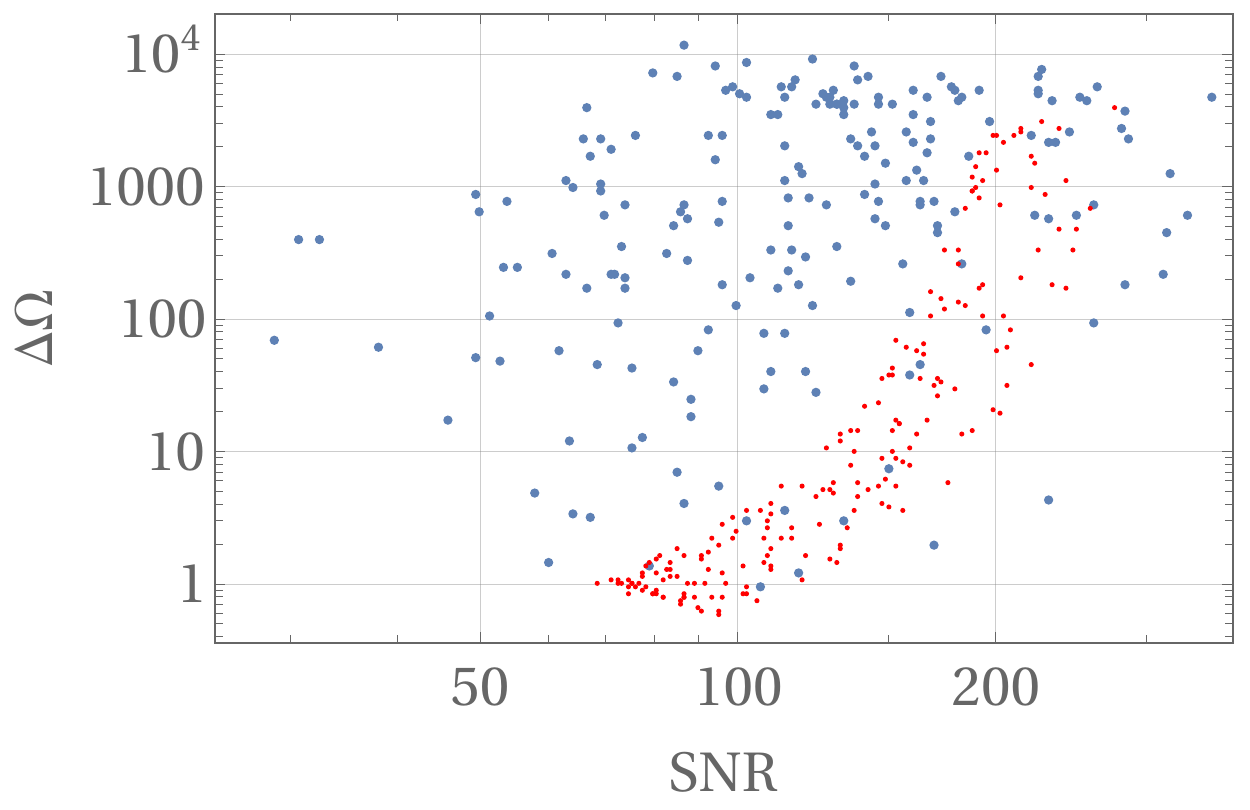}
	\end{minipage}
	\hspace{5mm}
	\begin{minipage}{0.4\linewidth}
		\centering
		\includegraphics[width=\linewidth]{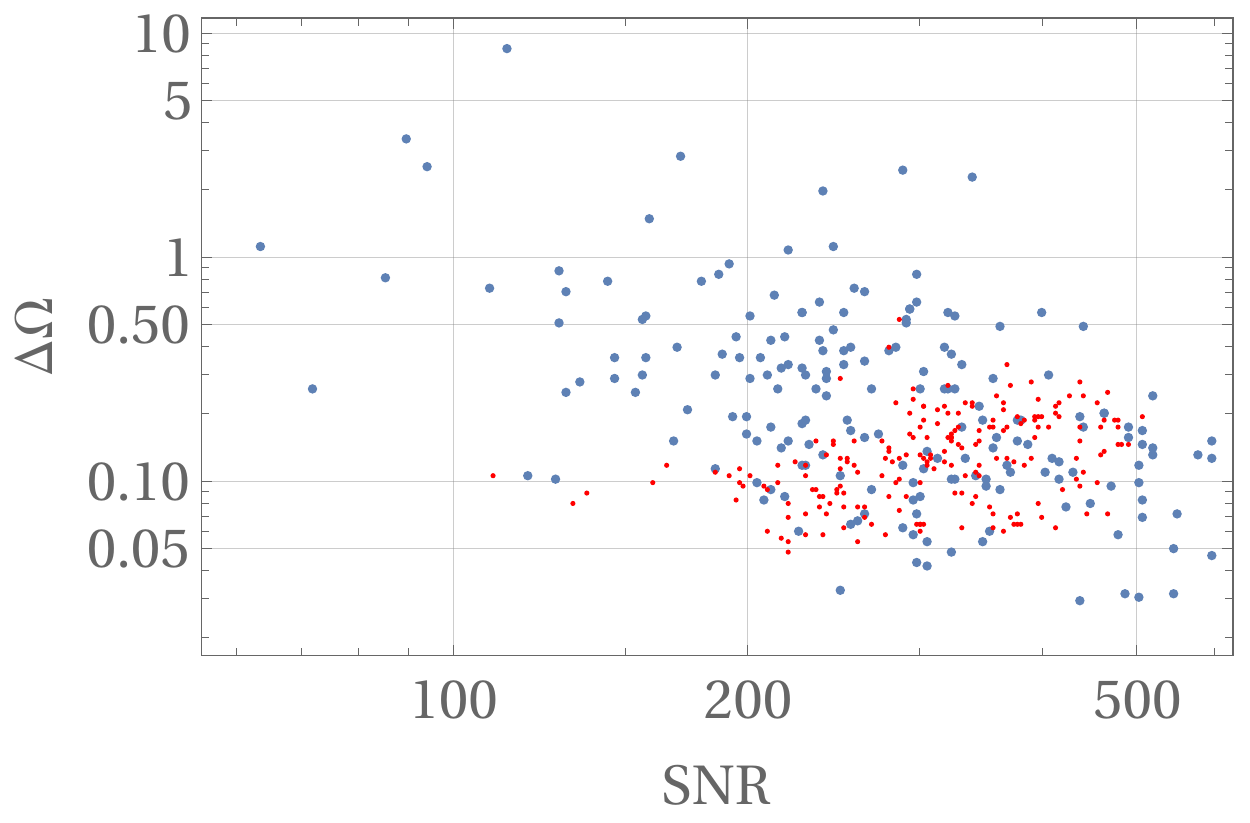}
	\end{minipage}
	\end{tabular}
	\caption{SNR dependences of sky localization errors for NSBH with $m_1 / m_2 = 10$.  Left panel is for single detector case, whereas Right panel is for double detectors case.  Blue (Red) dots are with the earth rotation without (with) precession effect.}
	\label{FIG:result_NSBH_SNRscaling}
\end{figure*}

However, for sources at a fixed redshift but with different precession parameters, the SNR scalings are nontrivial and do not hold in the single detector case.
The estimated errors for NSBH with $m_1 / m_2 = 10$ are shown as a function of SNR in \figref{FIG:result_NSBH_SNRscaling}.
The sky localization errors in the single detector case do not simply decrease as the SNR increases.
For a non-precessing case, the errors are highly scattered, indicating that the degeneracies between the extrinsic parameters exist.
For a precessing case, the localization errors scale with SNR even inversely.
These can be understood as follows.
GW amplitude is proportional to the norm of Eq.~\eqref{EQ:z}, which oscillates by the change of $\alpha$.
Neglecting the oscillation, that is, setting $\alpha(v) = 0$, the norm is smoothly changed by $\beta$\footnote{
	If $\alpha(v)=\pi$, the phase is $\beta + \theta_J$ instead of $\beta - \theta_J$.
	The phase in the cosine functions is the angle between $\vec{L}$ and the line of sight.
}:
\eq{
   \left. |Z|^2 \right|_{\alpha=0} = F_\cross^2 \cos^2(\beta - \theta_J) + \frac{1}{16} F_+^2 \{ 3 + \cos2(\beta - \theta_J) \}^2 \;,
}
which is plotted as a function of $\beta - \theta_J$ in \figref{FIG:g2}.
Since $\beta(v)$ increases as time goes, $|Z|^2$ decreases if $\beta - \theta_J$ starts from $\sim 0$ or $\pi$ at low frequencies and increases if it starts from $\sim \pi/2$.
As SNR is more earned at lower frequencies and the localization errors, $\varDelta\Omega$ and $\varDelta d_L / d_L$, are mostly determined at higher frequencies, where GW amplitude is larger and a timing error is $\sim 1/f$, high (low) SNR events have smaller (larger) GW amplitude at high frequencies and give worse (better) localization errors.
Thus, the errors of the extrinsic parameters increase with respect to SNR.
The physical interpretation for the inverse SNR scalings is that tilting $\vec{L}$ from $\vec{J}$ as time evolves, and then the flux of {GW} to observer increase or decrease depending on the initial inclination.
Thus the inverse SNR scaling is just a bias when a redshift is fixed.

The double detector case obeys a normal scaling with SNR, because the degeneracies between the inclination and other extrinsic parameters are almost broken by measuring {GW}s with different orientating detectors.

\myfig[width=0.8\linewidth]{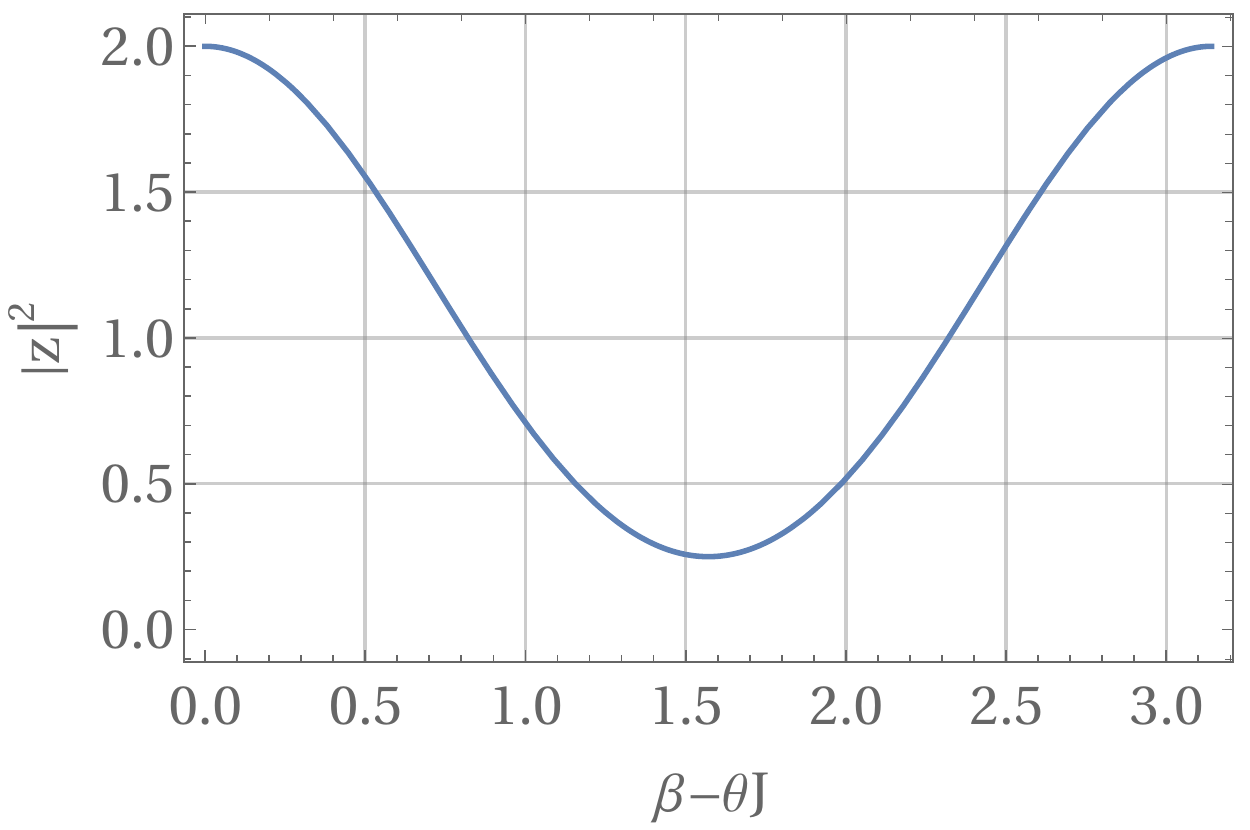}{$\beta - \theta_J$ vs $|Z|^2$ for $\alpha(v) = 0, F_\cross = 0, F_+ = 1$.}{FIG:g2}


\bibliographystyle{unsrt}
\bibliography{references}

\end{document}